\documentclass[aps,pra,twocolumn,preprintnumbers,amsmath,amssymb,superscriptaddress]{revtex4-2}

\usepackage{subfigure}
\usepackage{graphicx,amsmath}
\usepackage{epsf}
\usepackage[colorlinks = true, linkcolor = blue, urlcolor  = blue, citecolor = blue, anchorcolor = blue]{hyperref}

\usepackage{bm}
\usepackage{dsfont}
\usepackage{natbib}

\begin{document}
\title{Slow Light through Brillouin Scattering in Continuum Quantum Optomechanics}
\author{Hashem~Zoubi}
\email{hashemz@hit.ac.il}
\affiliation{Department of Physics, Holon Institute of Technology, Holon 5810201, Israel}
\author{Klemens~Hammerer}
\affiliation{Institute for Theoretical Physics, Leibniz University Hanover, Appelstrasse 2, 30167 Hanover, Germany}
\date{24 May 2024}

\begin{abstract} 
We investigate the possibility of achieving a slow signal field at the level of single photons inside nanofibers by exploiting stimulated Brillouin scattering, which involves a strong pump field and the vibrational modes of the waveguide. The slow signal is significantly amplified for a pump field with a frequency higher than that of the signal, and attenuated for a lower pump frequency. We introduce a configuration for obtaining a propagating slow signal without gain or loss and with a relatively wide bandwidth. This process involves two strong pump fields with frequencies both higher and lower than that of the signal, where the effects of signal amplification and attenuation compensate each other. We account for thermal fluctuations due to the scattering off thermal phonons and identify conditions under which thermal contributions to the signal field are negligible. The slowing of light through Brillouin optomechanics may serve as a vital tool for optical quantum information processing and quantum communications within nanophotonic structures.
\end{abstract}

\maketitle

\section{Introduction}

In recent years, significant progress has been achieved in fabricating waveguides with cross-sections nearing nanoscale dimensions~\cite{Safavi2019}, opening new horizons for Stimulated Brillouin Scattering (SBS). A pivotal advancement in SBS emerged with the identification of a dominant mechanism induced by radiation pressure, as theoretically predicted by \cite{Rakich2012,Rakich2018,VanLaer2016,Zoubi2016} and experimentally realized by \cite{Shin2013,Beugnot2014,VanLaer2015a,VanLaer2015b,Kittlaus2016,Kittlaus2017}. SBS in waveguides has found application across a broad spectrum of communication and information processing technologies. The substantial enhancement of SBS in waveguides facilitates the amplification of the Stokes field \cite{Kittlaus2016,Kittlaus2017,Otterstrom2019}, paving the path towards a Brillouin laser \cite{Otterstrom2018a,Gundavarapu2019,Chauhan2021} and light storage \cite{Zhu2007,Merklein2017}. Various proposals for nanoscale waveguides \cite{Safavi2019} have emerged in the literature, where the waveguide's mechanical quality factor—determining the sound wave lifetime \cite{Eggleton2013}—significantly impacts the efficiency of each proposed device. Furthermore, thermal phonons pose major challenges to efficient photon and phonon processes within waveguides \cite{VanLaer2017,Kharel2016,Behunin2018,Dallyn2022}. To address these challenges, optomechanical cooling via sideband cooling in a continuous system was demonstrated, using SBS to cool a continuum of traveling wave phonons in a waveguide by tens of Kelvins \cite{Otterstrom2018b}. These achievements open up the possibility to develop versatile light-matter interfaces~\cite{Hammerer2010} based on SBS achieving, for example, optomechanical entanglement~\cite{Zhu2024} or nonlinear photon interactions~\cite{Zoubi2017}.

In this paper, we introduce a configuration to achieve slow photons using SBS within waveguides. By coupling a signal field to classical pump fields through Brillouin scattering mediated by acoustic waves it is possible to achieve a low effective group velocity; however, the signal's amplitude is significantly amplified when the pump frequency exceeds that of the signal and considerably attenuated when the pump frequency is lower \cite{Thevenaz2008}. A stable signal amplitude can be maintained by employing two simultaneous pump fields with frequencies both above and below that of the signal. The Brillouin scattering from the higher pump field into the signal is balanced by the scattering from the signal field into the lower pump field. We account for the impact of thermal phonons in the waveguide medium and identify conditions under which thermal contributions to the signal amplitude are negligible. The real-space quantum Langevin equations of motion for the signal field are solved by assuming classical pump fields and adiabatically eliminating the phonon components. As a result, the signal field propagates through the waveguide without any gain or loss, with an effective group velocity significantly reduced compared to the group velocity of light. 

The ability to control the group velocity of light within waveguides opens new avenues for enhancing light-matter interactions, which are crucial for optical quantum information processing \cite{Obrien2007}. By slowing down the photons, their interaction time with the medium is extended, potentially increasing the efficiency of quantum gates and other processing elements \cite{Zoubi2017,Zoubi2021,Zoubi2023}. Additionally, the stable propagation of slow light without gain or loss is essential for maintaining the coherence of quantum states necessary for quantum communication and computing.

The paper is structured as follows: Section~\ref{Sec:ContinuumOptomechanics} introduces a coupled system of photons and phonons via SBS within a waveguide. Section~\ref{Sec:SlowLight} describes two methodologies to achieve a slow propagating signal field utilizing SBS and a strong classical pump field. The first method employs a pump field with a frequency higher than that of the signal, leading to significant signal amplification. The second method utilizes a pump field with a frequency lower than the signal's, resulting in considerable signal attenuation. In both scenarios, the impact of thermal fluctuations is analyzed. Section~\ref{Sec:SlowLightNoGain} discusses achieving a slow signal at the single-photon level without gain or loss by implementing two pump fields with frequencies both above and below that of the signal, while minimizing thermal contributions. Section~\ref{Sec:Conclusion} provides discussions and conclusions. Detailed derivations of the equations of motion and their solutions are presented in the appendices.

\section{Continuum quantum optomechanics in nanophotonic wires}\label{Sec:ContinuumOptomechanics}

\begin{figure}[t]
\includegraphics[width=1\linewidth]{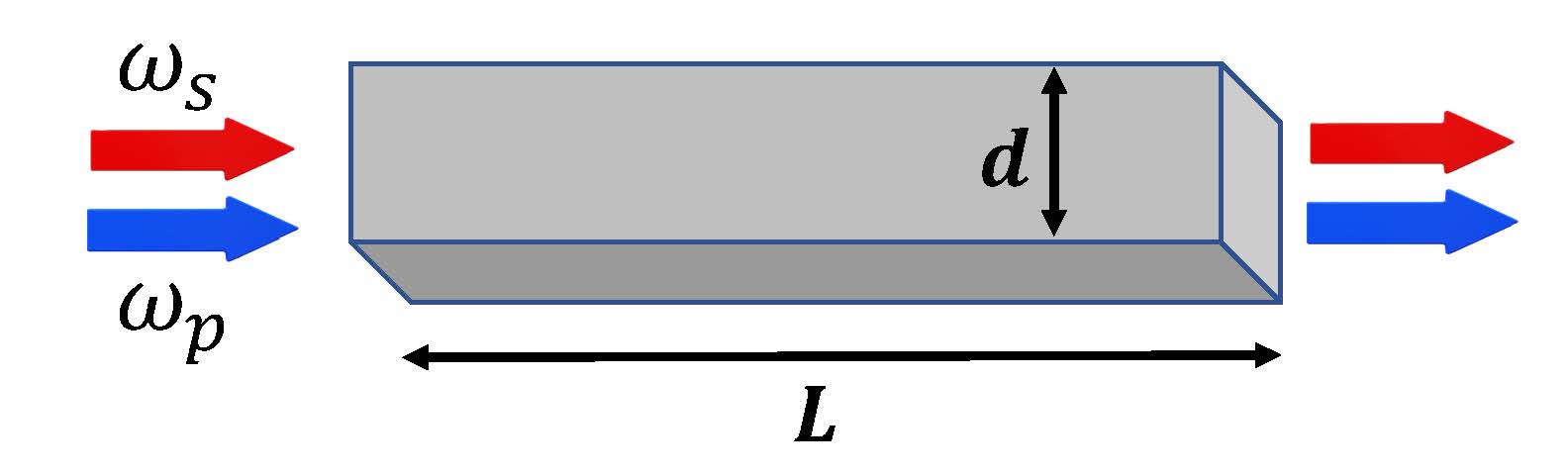}
\caption{A schematic diagram of a waveguide of length $L$ and dimension $d$, where $L\gg d$. The input-output pump and signal fields are presented at the two waveguide edges. The light wavelength $\lambda$ obeys $\lambda\lesssim d$.}
 \label{Fig1}
\end{figure}

We start by presenting a system of interacting light and sound waves within nanoscale waveguides via Brillouin scattering. The system consists of a waveguide composed of dielectric material placed in free space, characterized by a refractive index $n$ greater than one (e.g., for silicon material, $n\approx 3.5$), as depicted in figure (\ref{Fig1}). The length of the waveguide, $L$, significantly exceeds its transverse dimension, $d$, with $L\gg d$, and the light wavelength $\lambda$ is comparable to the wire dimension, $\lambda\lesssim d$. In our prior research \cite{Zoubi2016}, we formulated a microscopic quantum theory for the interaction between the light field and mechanical excitations in nanoscale waveguides, deriving a Brillouin-type Hamiltonian for the interplay of photons and phonons. This configuration allows photons and phonons to propagate freely along the waveguide while being confined in the transverse direction, leading to the emergence of photonic and phononic multi-mode branches. In \cite{Zoubi2016} we derived the dispersion realtions for photons and phonons and determined the photon-phonon coupling parameter by considering both electrostriction and radiation pressure mechanisms. In such an environment, the coupling of photon-phonon via Brillouin scattering is significantly intensified compared to conventional waveguides, a phenomenon corroborated by experimental findings \cite{Rakich2012,Shin2013,Kharel2016,VanLaer2016,Zoubi2016,Beugnot2014,VanLaer2015a,VanLaer2015b,Kittlaus2016,Kittlaus2017}. This research broadens the scope of conventional quantum optomechanics, which typically focuses on localized modes of photons and phonons, to include continuum quantum optomechanics encompassing propagating modes \cite{Aspelmeyer2014,Zoubi2016,Kharel2016,Safavi2019}.

The Hamiltonian for propagating photons within a waveguide is described by
\begin{equation}
H_{\mathrm{phot}}=\sum_{k,\alpha}\hbar\omega_{k\alpha}\ \hat{a}_{k\alpha}^{\dagger}\hat{a}_{k\alpha},
\end{equation}
where $\hat{a}_{k\alpha}^{\dagger}$ and $\hat{a}_{k\alpha}$ represent the creation and annihilation operators for a photon of wavenumber $k$ and branch $\alpha$, respectively, with $\omega_{k\alpha}$ denoting the photon frequency. The Hamiltonian for propagating phonons within a waveguide is expressed as
\begin{equation}
H_{\mathrm{phon}}=\sum_{q,\mu}\hbar\Omega_{q\mu}\ \hat{b}_{q\mu}^{\dagger}\hat{b}_{q\mu},
\end{equation}
where $\hat{b}_{q\mu}^{\dagger}$ and $\hat{b}_{q\mu}$ are the creation and annihilation operators for a phonon of wavenumber $q$ and branch $\mu$, respectively, with $\Omega_{q\mu}$ indicating the phonon frequency. The Hamiltonian describing the photon-phonon interaction is given by
\begin{equation}
H_{\mathrm{phot-phon}}=\hbar\sum_{k,q}\sum_{\alpha,\beta,\mu}\left\{g_{kq,\alpha\beta\mu}^{\ast}\ \hat{b}_{q\mu}^{\dagger}\hat{a}_{k-q\beta}^{\dagger}\hat{a}_{k\alpha}+\mathrm{h.c.}
\right\},
\end{equation}
where $g_{kq,\alpha\beta\mu}$ represents the photon-phonon coupling parameter. The first term describes the scattering of a photon from wavenumber $k$ in branch $\alpha$ to wavenumber $k-q$ in branch $\beta$ through the emission of a phonon of wavenumber $q$ in branch $\mu$. Conversely, the Hermitean conjugate (h.c.) term accounts for scattering of a photon from wavenumber $k-q$ in branch $\beta$ to wavenumber $k$ in branch $\alpha$ via the absorption of a phonon of wavenumber $q$ in branch $\mu$. Owing to translational symmetry along the wire, these processes adhere to momentum conservation. Note that the momentum-space operators for photons and phonons, $\hat{a}_{k\alpha}$ and $\hat{b}_{q\mu}$, are dimensionless.

In \cite{Zoubi2016}, we solved the equations of motion for the electromagnetic field and mechanical excitation to derive the photon and phonon dispersions analytically for the specific case of a cylindrical waveguide, obtaining the frequencies $\omega_{k\alpha}$ and $\Omega_{q\mu}$. However, the scheme of the current paper can be implemented experimentally for nanoscale wires of any cross-section shape, e.g. circular and rectangular \cite{Rakich2012,Kittlaus2017,Safavi2019}. We focus here on a linear region of the dispersion and assume that the light injected into the waveguide possesses a finite bandwidth. For photons, we employ the linear dispersion relation $\omega_{k\alpha} = \omega_{0\alpha} + v_{g\alpha}(k - k_{0\alpha})$, where $\omega_{0\alpha}$ is the frequency at the center of the signal bandwidth for branch $\alpha$. The wavenumber bandwidth is denoted by ${\cal B}_{0\alpha}^k$ around $k_{0\alpha}$. The effective group velocity in the linear segment is $v_{g\alpha}$ for branch $\alpha$. A similar approach is applied to the phonon dispersion, where $\Omega_{q\mu} = \Omega_{0\mu} + v_{s\mu}(q - q_{0\mu})$. The wavenumber bandwidth ${\cal B}_{0\mu}^q$ is centered around $q_{0\mu}$, with the sound velocity being $v_{s\mu}$ for branch $\mu$. For both propagating photons and phonons, the wavenumbers are determined by the periodic boundary condition in a wire of length $L$, where the wavenumber is quantized as $k = \frac{2\pi}{L}m$ with $m$ being integers $(m = 0, \pm1, \pm2, \cdots)$.
We convert the Hamiltonian from momentum-space to real-space representation to accommodate the space-time dynamics of pulse light fields propagating through the waveguide. This transformation is achieved by defining the light field operator as
\begin{equation}
\hat{\psi}_{\alpha}(z)=\frac{1}{\sqrt{L}}\sum_{k\in{\cal B}_{0\alpha}^k}\hat{a}_{k\alpha}e^{i(k-k_{0\alpha})z},
\end{equation}
and its inverse transformation by
\begin{equation}
\hat{a}_{k\alpha}=\frac{1}{\sqrt{L}}\int_0^L dz\ \hat{\psi}_{\alpha}(z)e^{-i(k-k_{0\alpha})z}.
\end{equation}
Translational symmetry ensures the identities $\frac{1}{L}\sum_{k}e^{-ik(z-z')}=\delta(z-z')$ and $\frac{1}{L}\int_0^L dz e^{i(k-k')z}=\delta_{k,k'}$, allowing field operators to satisfy the boson commutation relations $[\hat{\psi}_{\alpha}(z),\hat{\psi}_{\alpha}^{\dagger}(z')]=\delta(z-z')$. The real-space photon Hamiltonian is expressed as
\begin{multline}
H_{\mathrm{phot}}=\sum_{\alpha}\left\{\hbar\omega_{0\alpha}\int dz\ \hat{\psi}_{\alpha}^{\dagger}(z)\hat{\psi}_{\alpha}(z)\right.\\
\left.-i\hbar v_{g\alpha}\int dz\ \hat{\psi}_{\alpha}^{\dagger}(z)\frac{\partial\hat{\psi}_{\alpha}(z)}{\partial z}\right\}.
\end{multline}
This formulation allows for a nuanced treatment of the propagation dynamics of light pulses within the waveguide, encapsulating the effects of group velocity and phase shifts in real space.

Similarly, we define the mechanical excitation field operator as
\begin{equation}
\hat{\cal Q}_{\mu}(z)=\frac{1}{\sqrt{L}}\sum_{q\in{\cal B}_{0\mu}^q}\hat{b}_{q\mu}e^{i(q-q_{0\mu})z},
\end{equation}
and its inverse transformation by
\begin{equation}
\hat{b}_{q\mu}=\frac{1}{\sqrt{L}}\int_0^L dz\ \hat{\cal Q}_{\mu}(z)e^{-i(q-q_{0\mu})z},
\end{equation}
which satisfy the commutation relation $[\hat{\cal Q}_{\mu}(z),\hat{\cal Q}_{\mu}^{\dagger}(z')]=\delta(z-z')$. The real-space phonon Hamiltonian is given by
\begin{multline}
H_{\mathrm{phon}}=\sum_{\mu}\left\{\hbar\Omega_{0\mu}\int dz\ \hat{\cal Q}_{\mu}^{\dagger}(z)\hat{\cal Q}_{\mu}(z)\right.\\
\left.-i\hbar v_{s\mu}\int dz\ \hat{\cal Q}_{\mu}^{\dagger}(z)\frac{\partial\hat{\cal Q}_{\mu}(z)}{\partial z}\right\}. 
\end{multline}
The real-space photon and phonon field operators, having a dimension of $1/\sqrt{\text{length}}$, represent slowly varying spatial amplitudes.

The coupling parameter for photon-phonon interaction, $g_{kq,\alpha\beta\mu}$, is considered constant across the photon and phonon bandwidths, ${\cal B}_{0\alpha}^k$ and ${\cal B}_{0\mu}^q$. Utilizing the local field approximation, the coupling parameter simplifies to $g_{\alpha\beta\mu}$. Consequently, the real-space photon-phonon interaction Hamiltonian is expressed as
\begin{multline}
H_{\mathrm{phot-phon}}=\hbar\sqrt{L}\sum_{\alpha,\beta,\mu}\int dz\ \left\{g_{\alpha\beta\mu}^{\ast}\ \hat{\cal Q}_{\mu}^{\dagger}(z)\hat{\psi}_{\beta}^{\dagger}(z)\hat{\psi}_{\alpha}(z)\right.\\\left.+g_{\alpha\beta\mu}\ \hat{\psi}_{\alpha}^{\dagger}(z)\hat{\psi}_{\beta}(z)\hat{\cal Q}_{\mu}(z)\right\}.
\end{multline}
This formulation provides a compact description of the interaction between photons and phonons in the real-space framework, accommodating the direct and inverse scattering processes.

\section{Slow light in Brillouin quantum optomechanics}\label{Sec:SlowLight}

\begin{figure}[t]
\includegraphics[width=1\linewidth]{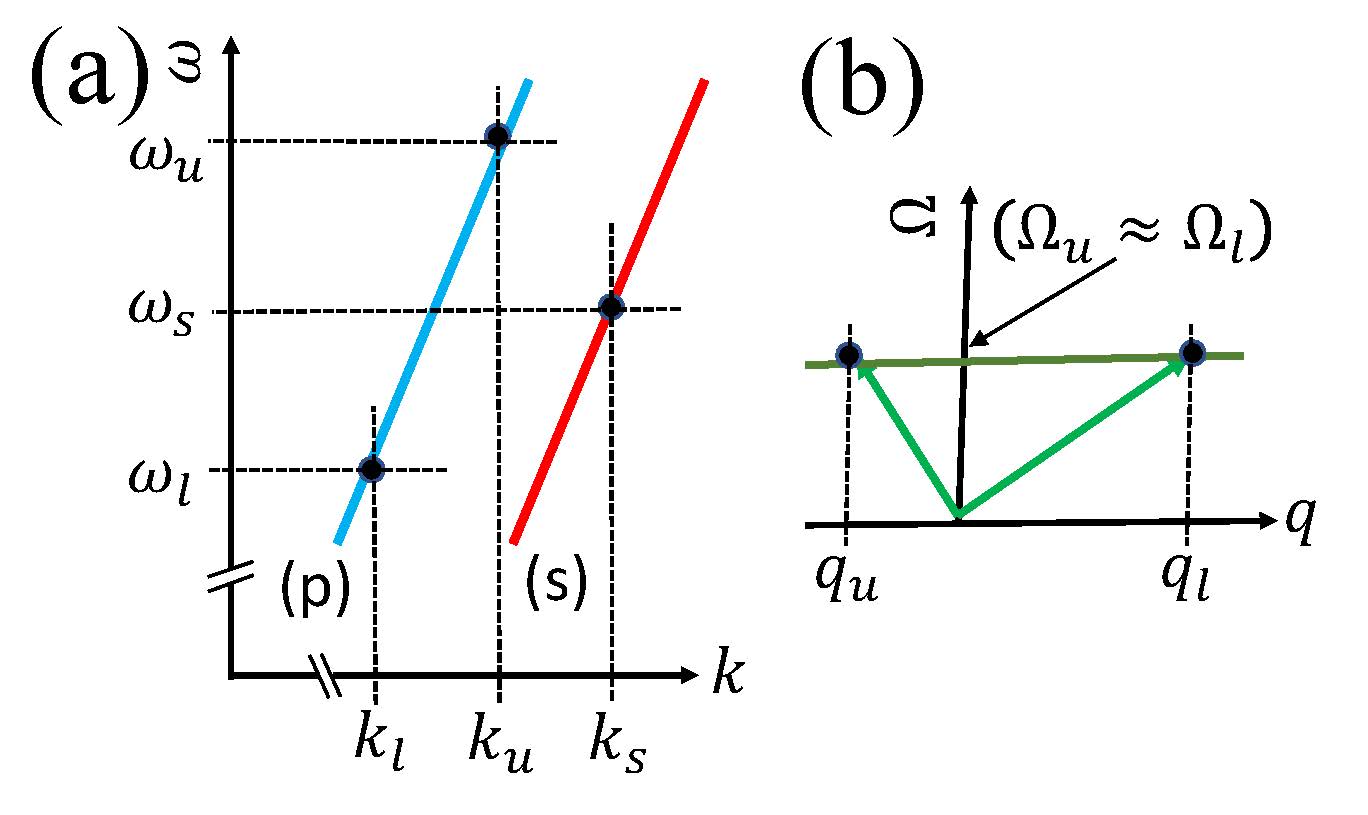}
\caption{(a) The photonic branches $(s)$ and $(p)$ are presented for the angular frequency $\omega$ as a function of the wavenumber $k$. The two branches are assumed to have linear dispersion in the appropriate zones with the same group velocity $v_g$. The relevant photon modes treated in the paper are indicated, which are the two pump fields $(\omega_u,k_u)$ and $(\omega_l,k_l)$, and the signal field $(\omega_s,k_s)$. (b) The phononic branch is presented for the angular frequency $\Omega$ as a function of the wavenumber $q$. The branch is assumed to be dispersionless in the appropriate zone. The relevant phonon modes treated in the paper are indicated, which are $(\Omega_u,q_u)$ and $(\Omega_l,q_l)$, where $\Omega_u=\Omega_l$ with $q_u\neq q_l$.}
 \label{Fig2}
\end{figure}

We proceed to describe on this basis the phenomena of signal field amplification and attenuation by exploiting stimulated inter-modal Brillouin scattering of co-propagating photons that belong to distinct
spatial optical modes \cite{Kittlaus2017}. We assume a signal field in branch $(s)$, centered around frequency $\omega_{0s}=\omega_s$. Conversely, the pump field occupies a distinct branch $(p)$ and is centered around frequency $\omega_{0p}=\omega_p$. Both branches are assumed to share identical slopes, leading to equal group velocities $v_g$ for the fields in each branch, as depicted in figure (\ref{Fig2}.a). The pump field, being considerably stronger than the signal field, is treated as a classical quantity with a stationary (slowly varying) amplitude denoted by ${\cal E}=\langle\hat{\psi}_p\rangle$. On the phonon side, we assume a non-dispersive single branch with a constant frequency $\Omega_{q\mu}=\Omega_{0\mu}=\Omega$, and negligible sound velocity $v_{s\mu}$, as shown in figure (\ref{Fig2}.b). Consequently, the photon Hamiltonian is formulated as
\begin{equation}\label{SigHam}
H_{\mathrm{phot}}=\hbar\omega_{s}\int dz\ \hat{\psi}_{s}^{\dagger}(z)\hat{\psi}_{s}(z)-i\hbar v_{g}\int dz\ \hat{\psi}_{s}^{\dagger}(z)\frac{\partial\hat{\psi}_{s}(z)}{\partial z}.
\end{equation}
Both the signal and pump fields are assumed to propagate in the rightward direction. The rate of photon damping is considered negligible during their transit along the waveguide's length $L$. Phonon dissipation is accounted for by incorporating a damping rate $\Gamma$, and thermal fluctuations are represented through the Langevin force operators $\hat{\cal F}$, adhering to the properties outlined in \cite{Gardiner2010}
\begin{eqnarray}\label{LF}
\langle\hat{\cal F}(z,t)\hat{\cal F}(z',t')\rangle&=&\langle\hat{\cal F}^{\dagger}(z,t)\hat{\cal F}^{\dagger}(z',t')\rangle=0, \nonumber \\
\langle\hat{\cal F}^{\dagger}(z,t)\hat{\cal F}(z',t')\rangle&=&\Gamma \bar{n}\ \delta(t-t')\delta(z-z'), \nonumber \\
\langle\hat{\cal F}(z,t)\hat{\cal F}^{\dagger}(z',t')\rangle&=&\Gamma (\bar{n}+1)\ \delta(t-t')\delta(z-z'),
\end{eqnarray}
with $\bar{n}$ representing the average phonon count at frequency $\Omega$. At low temperatures the appearance of thermal photons is negligible, while thermal phonons are likely present and treated here as a heat reservoir in applying the Markovian approximation \cite{Gardiner2010}.

In our analysis, we explore two distinct scenarios based on the relationship between the pump and signal frequencies: (1) the pump frequency is higher than that of the signal, denoted as $\omega_p>\omega_s$, and (2) the pump frequency is lower than the signal frequency, indicated by $\omega_p<\omega_s$.

\subsection{Slow light with signal amplification}

\begin{figure}[t]
\includegraphics[width=1\linewidth]{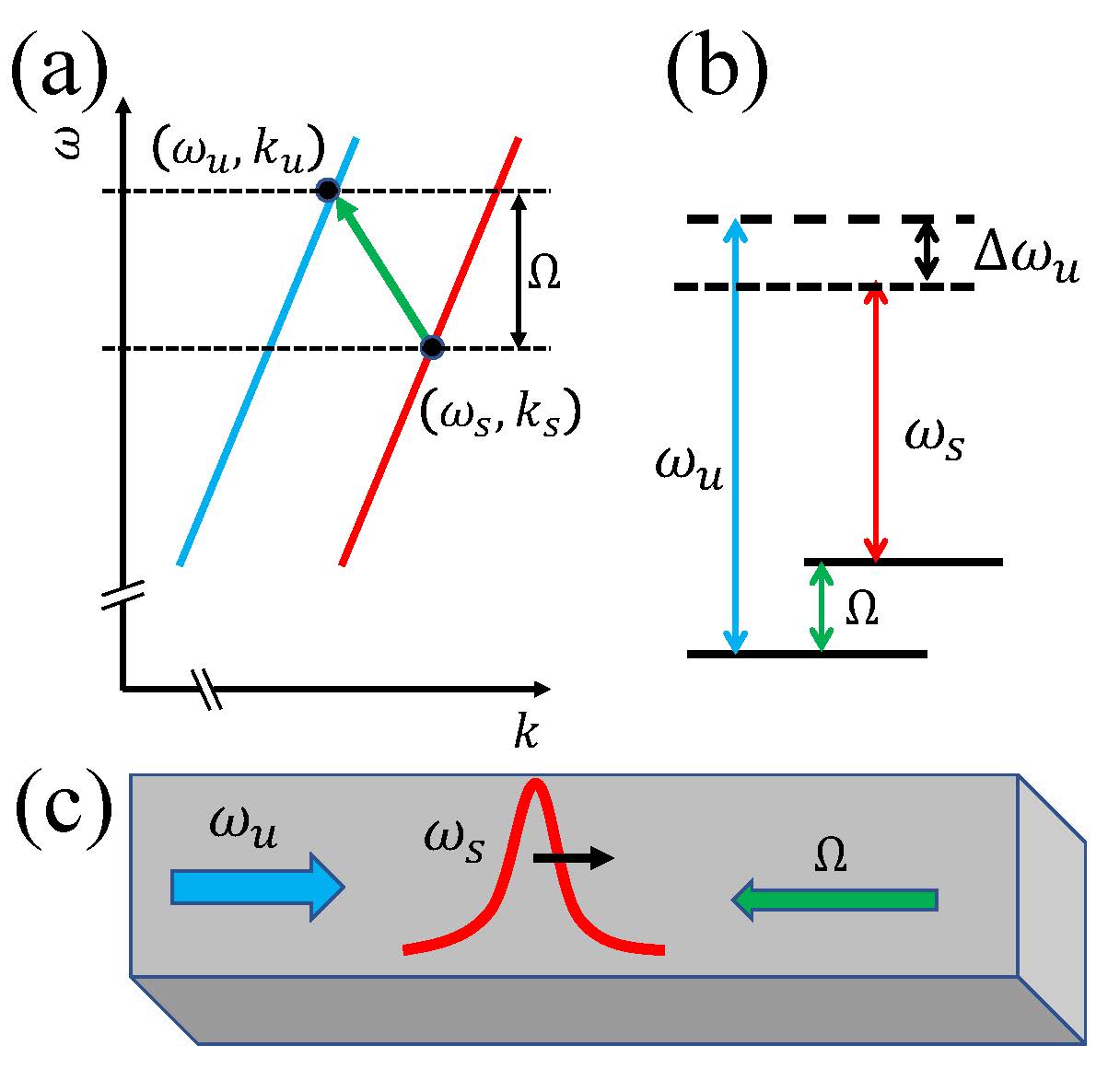}
\caption{(a) A pump field $(\omega_u,k_u)$ scatters into a signal field $(\omega_s,k_s)$ by the emission of a phonon $(\Omega,q_u)$. The process obeys conservation of energy $\omega_u\approx \omega_s+\Omega$ and conservation of momentum $k_u-k_s\approx q_u$. (b) A schematic energy diagram of the photon and phonon modes. A pump photon (of frequency $\omega_u$) is annihilated and a signal photon (of frequency $\omega_s$) and a phonon (of frequency $\Omega$) are created. The detuning of the process is $\Delta\omega_u=\omega_u-\omega_s-\Omega$. (c) A signal field of frequency $\omega_s$ is propagating to the right, with a co-propagating classical pump field of frequency $\omega_u$, where $\omega_u>\omega_s$. Due to stimulated Brillouin scattering a pump photon scatters into a signal phonon by the emission of a counter-propagating phonon of frequency $\Omega$.}
 \label{Fig3}
\end{figure}

In the scenario where $\omega_p>\omega_s$, a pump photon is scattered into a signal photon through the emission of a phonon, or conversely, a signal photon is converted into a pump photon by the absorption of a phonon, as illustrated in figure (\ref{Fig3}). The amplitude of the pump field is represented by ${\cal E}_u$, with its frequency designated as $\omega_u\equiv\omega_p$. The phonon operator is expressed by $\hat{\cal Q}_{u}$, and the associated Hamiltonian for the phonons is formulated as
\begin{equation}\label{phonu}
H_{\mathrm{phon}}^u=\hbar\Omega\int dz\ \hat{\cal Q}_{u}^{\dagger}(z)\hat{\cal Q}_{u}(z).
\end{equation}
The interaction Hamiltonian between photons and phonons is given by
\begin{equation}\label{photphonu}
H_{\mathrm{phot-phon}}^u=\hbar\sqrt{L}\int dz\ \left\{g_{u}^{\ast}{\cal E}_u\ \hat{\cal Q}_{u}^{\dagger}(z)\hat{\psi}_{s}^{\dagger}(z)+\mathrm{h.c.}
\right\}.
\end{equation}

The Heisenberg-Langevin equations of motion for the photon and phonon field operators are formulated as 
\begin{align}\label{eqphotphonu}
\left(\frac{\partial}{\partial t}+v_g\frac{\partial}{\partial z}\right)\hat{\psi}_{s}(z,t)&=-i\omega_s\ \hat{\psi}_{s}(z,t)-i\sqrt{L}g_{u}^{\ast}{\cal E}_u\ \hat{\cal Q}_{u}^{\dagger}(z,t), \nonumber \\
\left(\frac{\partial}{\partial t}+\frac{\Gamma}{2}\right)\hat{\cal Q}_{u}(z,t)&=-i\Omega\ \hat{\cal Q}_{u}(z,t)-i\sqrt{L}g_{u}{\cal E}_u\ \hat{\psi}_{s}^{\dagger}(z,t)\nonumber\\&\quad
-\hat{\cal F}(z,t).
\end{align}
The resulting signal field operator is (see appendix A for details)
\begin{align}
\hat{\psi}_{s}(z,t)&=\hat{\psi}_{s}^{in}(z-v_gt)e^{(G_u-i\kappa_u)z}
+i\frac{\sqrt{L}g_{u}^{\ast}{\cal E}_u}{v_g}e^{-i\Delta\omega_ut}\nonumber\\
&\times\int_0^t dt'\int_0^{z}dz'
\hat{\cal F}^{\dagger}(z',t')e^{-\frac{\Gamma}{2}(t-t')}e^{(G_u-i\kappa_u)(z-z')},
\end{align}
where $\hat{\psi}_{s}^{in}(z-v_gt)$ represents the incoming signal field operator, and $\Delta\omega_u=\omega_u-\omega_s-\Omega$ denotes the detuning frequency, as illustrated in figure (\ref{Fig3}.b). The gain parameter is defined as
\begin{equation}\label{Gu}
G_u=\frac{2|g_u|^2}{v_g\Gamma}\left\{\frac{{\cal I}_u}{1+\Delta_u^2}\right\},
\end{equation}
and the shift in wavenumber is given by
\begin{equation}
\kappa_u=\frac{2|g_u|^2}{\Gamma v_g}\frac{\Delta_u{\cal I}_u}{1+\Delta_u^2},
\end{equation}
with $\Delta_u=2\Delta\omega_u/\Gamma$ representing the scaled detuning, and ${\cal I}_u=L|{\cal E}_u|^2$ denoting the dimensionless pump intensity.

Using the relations (\ref{LF}), the average number of photons per unit length, or photon density, is calculated as
\begin{multline}
    \langle\hat{\psi}_{s}^{\dagger}(z,t)\hat{\psi}_{s}(z,t)\rangle=\langle\hat{\psi}_{s}^{in\dagger}(z-v_gt)\hat{\psi}_{s}^{in}(z-v_gt)\rangle e^{2G_uz}\\+{\cal N}_u(z,t),
\end{multline}
where ${\cal N}_u(z,t)$ represents the thermal contribution
\begin{equation}
{\cal N}_u(z,t)=-\frac{|g_u|^2L|{\cal E}_u|^2}{2G_uv_g^2}(\bar{n}+1)\left(1-e^{-\Gamma t}\right)\left(1-e^{2G_uz}\right).
\end{equation}
In this formulation, correlations between the Langevin force operators and the initial signal operator are disregarded. This approach focuses on the significant impact of the gain and thermal noise on the evolution of the photon density within the medium, illustrating how amplification and thermal effects contribute to the overall behavior of the signal.

The effective group velocity is defined by
\begin{equation}
\frac{1}{v_e^u}=\frac{1}{v_g}-\frac{\partial\kappa_u}{\partial\omega_s}.
\end{equation}
We obtain
\begin{equation}\label{veu}
\frac{v_e^u}{v_g}=\left(1+\frac{4|g_u|^2}{\Gamma^2}\left\{{\cal I}_u\frac{\left[1-\Delta_u^2\right]}{\left[1+\Delta_u^2\right]^2}\right\}\right)^{-1}.
\end{equation}
The rate of change of the gain $G_u$ with respect to the signal frequency is
\begin{equation}\label{dGu}
\frac{\partial G_u}{\partial\omega_s}=\frac{8|g_u|^2}{v_g\Gamma^2}\left\{{\cal I}_u\frac{\Delta_u}{\left[1+\Delta_u^2\right]^2}\right\}.
\end{equation}

Our primary goal is to achieve a slow propagating signal, aiming for $\frac{v_e^u}{v_g}\ll1$, while also preferring the signal to propagate without significant gain, hence $G_uL\ll1$. Additionally, it is crucial to minimize the impact of thermal phonons, ensuring that ${\cal N}_uL\ll 1$. While the condition for slow light can be met, this comes at the cost of high signal amplification and increased thermal fluctuations. E.g. choosing as a specific physical example $g_u=10^6$ Hz, $\Gamma=10^8$ Hz, ${\cal I}_u=\frac{1}{4}\times10^8$, $L=10^{-2}$ m, $v_g=10^8$ m/s, and $\Delta_u=\frac{1}{2}$, we find $\frac{v_e^u}{v_g}\approx 2\times 10^{-4}$, and $\frac{\partial G_u}{\partial\omega_s}\approx 0.64\times10^{-4}$ s/m. This results in slow light with a relatively large bandwidth, yet with a substantial gain factor of $G_uL\approx 40$. At a phonon frequency of $\Omega=50$ GHz, an average number of thermal quanta $\bar{n}\approx 0.0224$ is achievable at a temperature of $T\approx 0.1$ K$^{\circ}$. At the waveguide's output ($z=L$), in the high gain limit of $G_uL\gg 1$, the thermal contribution becomes significant, leading to ${\cal N}_{out}^uL\gg1$.

\begin{figure}[t]
\includegraphics[width=0.75\linewidth]{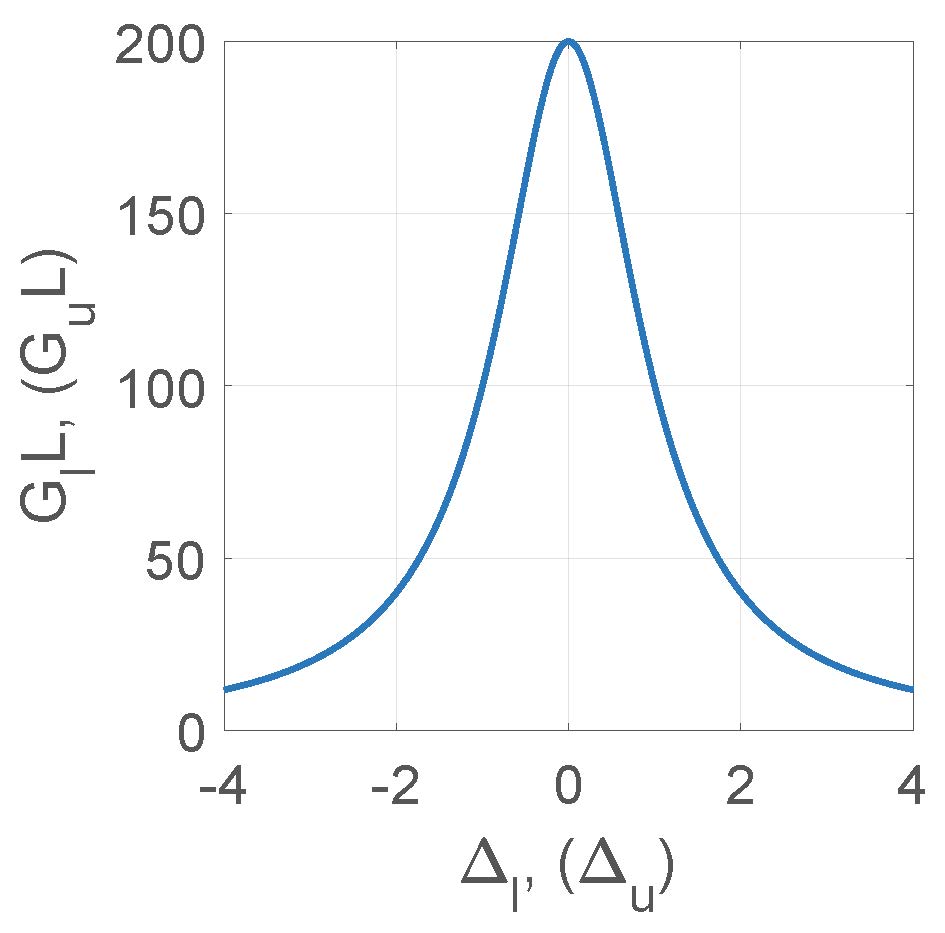}
\caption{The gain factor $G_uL\ (G_lL)$ as a function of the scaled detuning $\Delta_u\ (\Delta_l)$.}
 \label{Fig4}
\end{figure}

\begin{figure}[t]
\includegraphics[width=0.53\linewidth]{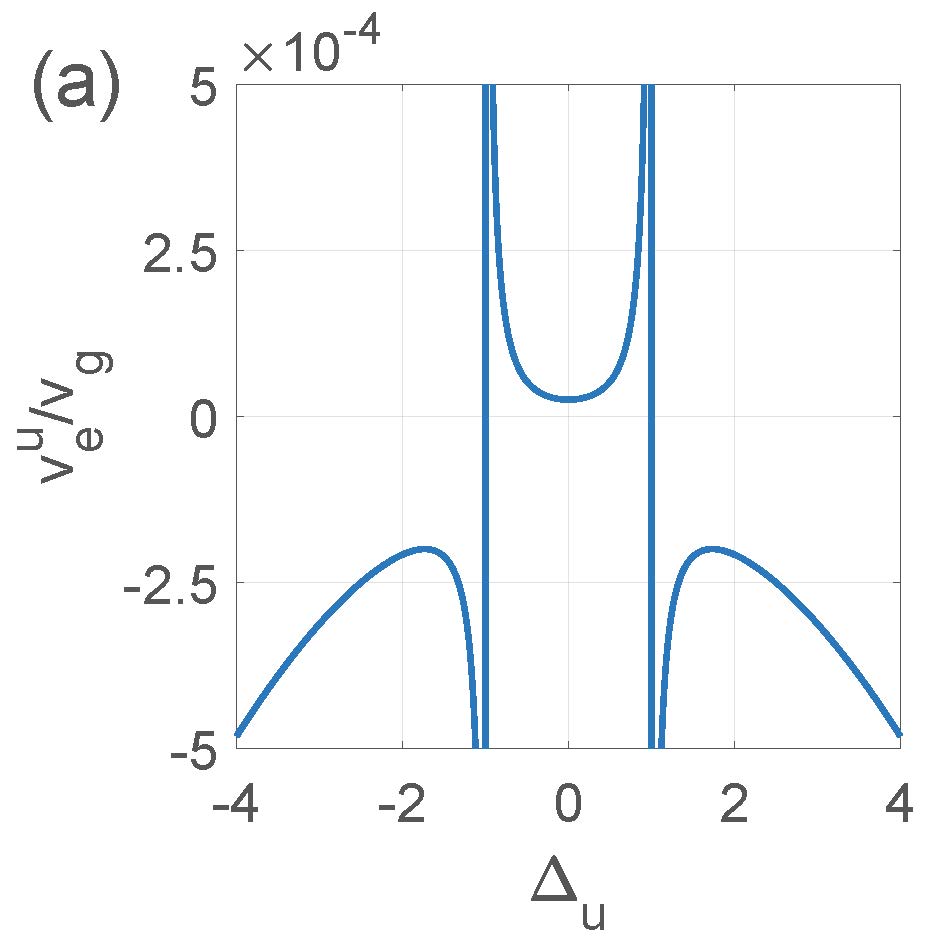}\includegraphics[width=0.5\linewidth]{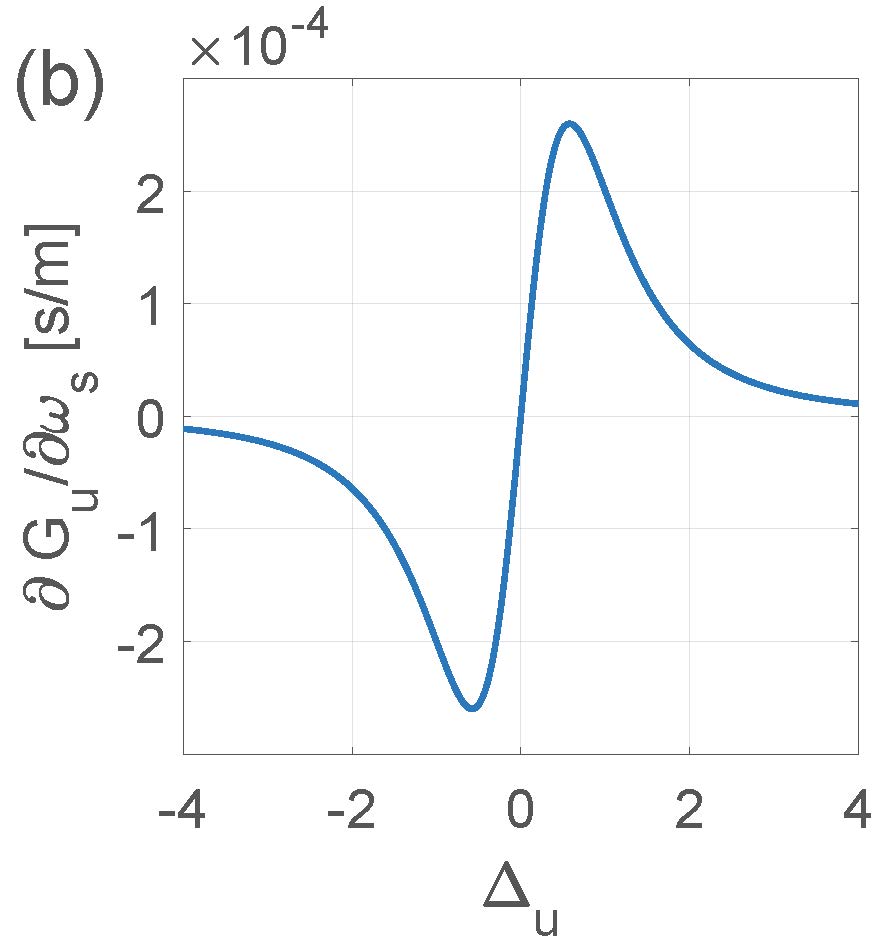}
\caption{(a) The relative effective group velocity $\frac{v_e^u}{v_g}$ as a function of the scaled detuning $\Delta_u$. (b) The rate of change of the gain factor with respect to the signal frequency $\frac{\partial G_u}{\partial\omega_s}$ as a function of the scaled detuning $\Delta_u$.}
 \label{Fig5}
\end{figure}

For the case of ${\cal I}_u=10^8$, in figure (\ref{Fig4}) we plot the gain factor $G_uL$ from equation (\ref{Gu}) as a function of $\Delta_u$. In figure (\ref{Fig5}.a), we plot the relative effective velocity $\frac{v_e^u}{v_g}$ from equation (\ref{veu}) as a function of $\Delta_u$. The rate of change of the gain factor with respect to the signal frequency, $\frac{\partial G_u}{\partial\omega_s}$ from equation (\ref{dGu}), is plotted in figure (\ref{Fig5}.b) as a function of $\Delta_u$. It is evident that the effective group velocity $v_e^u$ is significantly smaller than the group velocity $v_g$ around zero detuning, and the rate of change of the gain factor is negligible in the same zone. However, the gain factor $G_uL$ is large in this interval, leading to significant amplification of the signal photons.

\subsection{Slow light with signal attenuation}

\begin{figure}[t]
\includegraphics[width=1\linewidth]{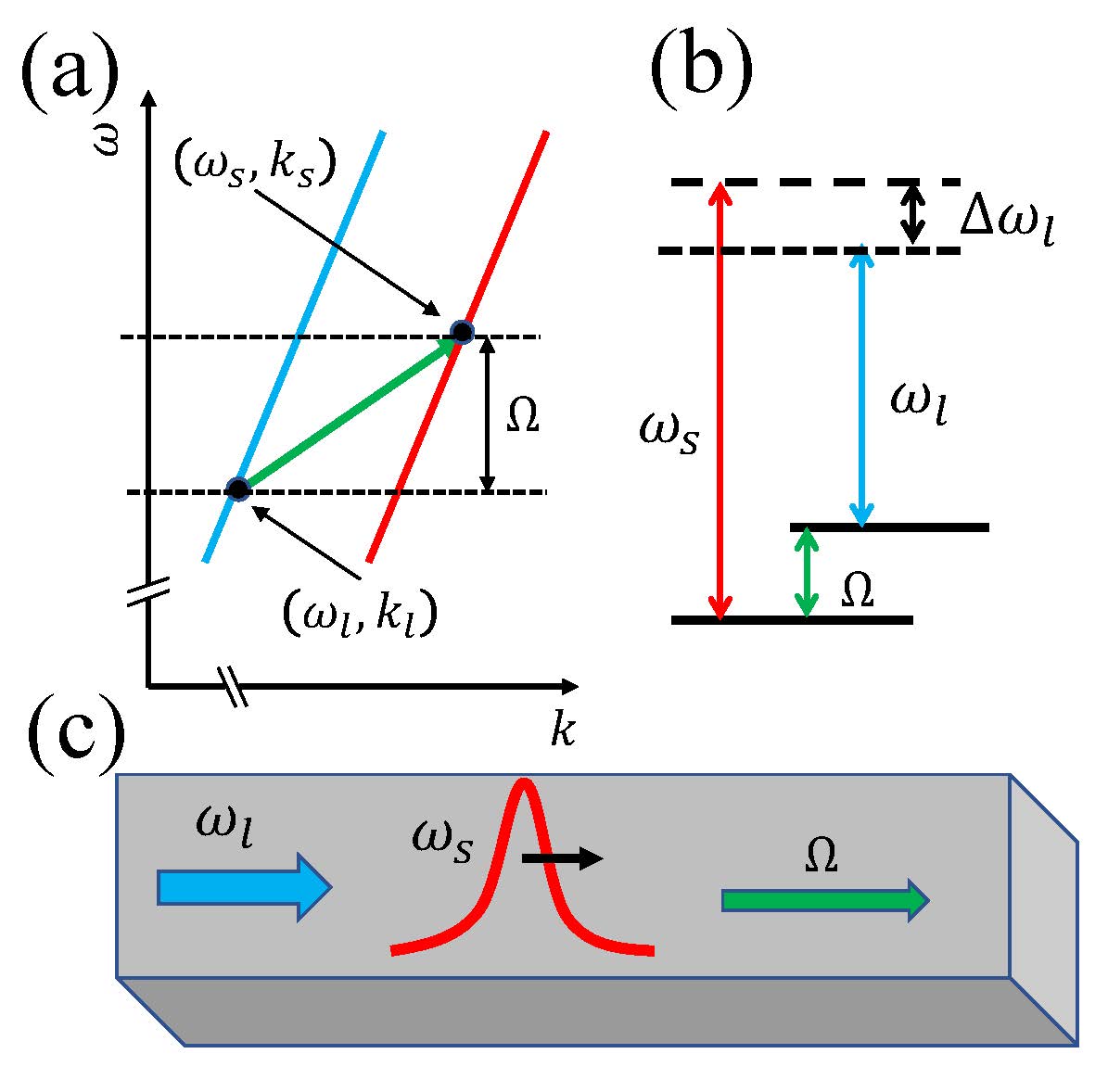}
\caption{(a) A signal field $(\omega_s,k_s)$ scatters into a pump field $(\omega_l,k_l)$ by the emission of a phonon $(\Omega,q_l)$. The process obeys conservation of energy $\omega_s\approx \omega_l+\Omega$ and conservation of momentum $k_s-k_l\approx q_u$. (b) A schematic energy diagram of the photon and phonon modes. A signal photon (of frequency $\omega_s$) is annihilated and a pump photon (of frequency $\omega_l$) and a phonon (of frequency $\Omega$) are created. The process detuning frequency is $\Delta\omega_l=\omega_s-\omega_l-\Omega$. (c) A signal field of frequency $\omega_s$ is propagating to the right, with a co-propagating classical pump field of frequency $\omega_l$, where $\omega_s>\omega_l$. Due to stimulated Brillouin scattering a signal photon scatters into a pump phonon by the emission of a co-propagating phonon of frequency $\Omega$.}
 \label{Fig6}
\end{figure}

For the scenario where $\omega_p<\omega_s$, a pump photon is scattered into a signal photon by the absorption of a phonon (or conversely, a signal photon is scattered into a pump photon by the emission of a phonon), as illustrated in figure (\ref{Fig6}). The amplitude of the pump field is represented by ${\cal E}_{l}$, with its frequency designated as $\omega_l\equiv\omega_p$. The phonon operator is denoted by $\hat{\cal Q}_{l}$, and the associated Hamiltonian for the phonons is formulated as
\begin{equation}\label{phonl}
H_{\mathrm{phon}}^l=\hbar\Omega\int dz\ \hat{\cal Q}_{l}^{\dagger}(z)\hat{\cal Q}_{l}(z).
\end{equation}
Furthermore, the interaction Hamiltonian between photons and phonons is described by
\begin{equation}\label{photphonl}
H_{\mathrm{phot-phon}}^l=\hbar\sqrt{L}\int dz\ \left\{g_{l}^{\ast}{\cal E}_l\ \hat{\cal Q}_{l}(z)\hat{\psi}_{s}^{\dagger}(z)+\mathrm{h.c.}
\right\}.
\end{equation}

The Heisenberg-Langevin equations of motion for the photon and phonon field operators are given by
\begin{align}\label{eqphotphonl}
\left(\frac{\partial}{\partial t}+v_g\frac{\partial}{\partial z}\right)\hat{\psi}_{s}(z,t)&=-i\omega_s\ \hat{\psi}_{s}(z,t)-i\sqrt{L}g_{l}^{\ast}{\cal E}_l\ \hat{\cal Q}_{l}(z,t), \nonumber \\
\left(\frac{\partial}{\partial t}+\frac{\Gamma}{2}\right)\hat{\cal Q}_{l}(z,t)&=-i\Omega\ \hat{\cal Q}_{l}(z,t)-i\sqrt{L}g_{l}{\cal E}_l^{\ast}\ \hat{\psi}_{s}(z,t)\nonumber\\
&\quad-\hat{\cal F}(z,t).
\end{align}
The solution to the equations of motion, as provided in appendix B, yields
\begin{align}
\hat{\psi}_{s}(z,t)&=\hat{\psi}_{s}^{in}(z-v_gt)e^{-(G_l+i\kappa_l)z}+i\frac{\sqrt{L}g_{l}^{\ast}{\cal E}_l}{v_g}e^{i\Delta\omega_lt}\nonumber\\
&\times\int_0^t dt'\int_0^{z}dz'\ \hat{\cal F}(z',t')e^{-\frac{\Gamma}{2}(t-t')}e^{-(G_l+i\kappa_l)(z-z')},
\end{align}
with the detuning frequency defined as $\Delta\omega_l=\omega_s-\omega_l-\Omega$, as illustrated in figure (\ref{Fig6}.b). The gain parameter is
\begin{equation}\label{Gu}
G_l=\frac{2|g_l|^2}{v_g\Gamma}\left\{\frac{{\cal I}_l}{1+\Delta_l^2}\right\},
\end{equation}
and the wavenumber shift is
\begin{equation}
\kappa_l=\frac{2|g_l|^2}{\Gamma v_g}\frac{\Delta_l{\cal I}_l}{1+\Delta_l^2},
\end{equation}
where $\Delta_l=2\Delta\omega_l/\Gamma$ represents the scaled detuning and ${\cal I}_l=L|{\cal E}_l|^2$ signifies the dimensionless pump intensity.

The photon density, using relations (\ref{LF}), is given by
\begin{multline}
\langle\hat{\psi}_{s}^{\dagger}(z,t)\hat{\psi}_{s}(z,t)\rangle=\langle\hat{\psi}_{s}^{in\dagger}(z-v_gt)\hat{\psi}_{s}^{in}(z-v_gt)\rangle e^{-2G_lz}\\+{\cal N}_l(z,t),
\end{multline}
where the thermal contribution is defined by
\begin{equation}
{\cal N}_l(z,t)=\frac{|g_l|^2L|{\cal E}_l|^2}{2G_lv_g^2}\bar{n}\left(1-e^{-\Gamma t}\right)\left(1-e^{-2G_lz}\right).
\end{equation}

The effective group velocity is given by
\begin{equation}
\frac{1}{v_e^l}=\frac{1}{v_g}-\frac{\partial\kappa_l}{\partial\omega_s}.
\end{equation}
This leads to
\begin{equation}\label{vel}
\frac{v_e^l}{v_g}=\left(1-\frac{4|g_l|^2}{\Gamma^2}\left\{{\cal I}_l\frac{\left[1-\Delta_l^2\right]}{\left[1+\Delta_l^2\right]^2}\right\}\right)^{-1}.
\end{equation}
The rate of change of the gain with respect to the signal frequency is calculated as
\begin{equation}\label{dGl}
\frac{\partial G_l}{\partial\omega_s}=-\frac{8|g_l|^2}{v_g\Gamma^2}\left\{{\cal I}_l\frac{\Delta_l}{\left[1+\Delta_l^2\right]^2}\right\}.
\end{equation}

Our main goal is to achieve a slow propagating signal, aiming for $\frac{v_e^l}{v_g}\ll1$. It is essential for the signal to propagate without loss along the wire, requiring $G_lL\ll1$. Additionally, minimizing the influence of thermal phonons is crucial, ensuring ${\cal N}_lL\ll1$. Although achieving slow light is possible, it comes at the cost of high signal attenuation. Using the previously mentioned physical values, with ${\cal I}_l=10^8$ and $\Delta_l=2$, we find $\frac{v_e^l}{v_g}\approx 2\times 10^{-4}$, and $\frac{\partial G_l}{\partial\omega_s}\approx -0.64\times10^{-4}$ s/m. This scenario yields slow light with a relatively large bandwidth but incurs a significant loss factor of $G_lL\approx 40$. At the waveguide output, i.e., at $z=L$, and under the condition of high loss $G_lL\gg 1$, the thermal contribution becomes negligible, where ${\cal N}_{out}^lL\ll1$.

\begin{figure}[t]
\includegraphics[width=0.53\linewidth]{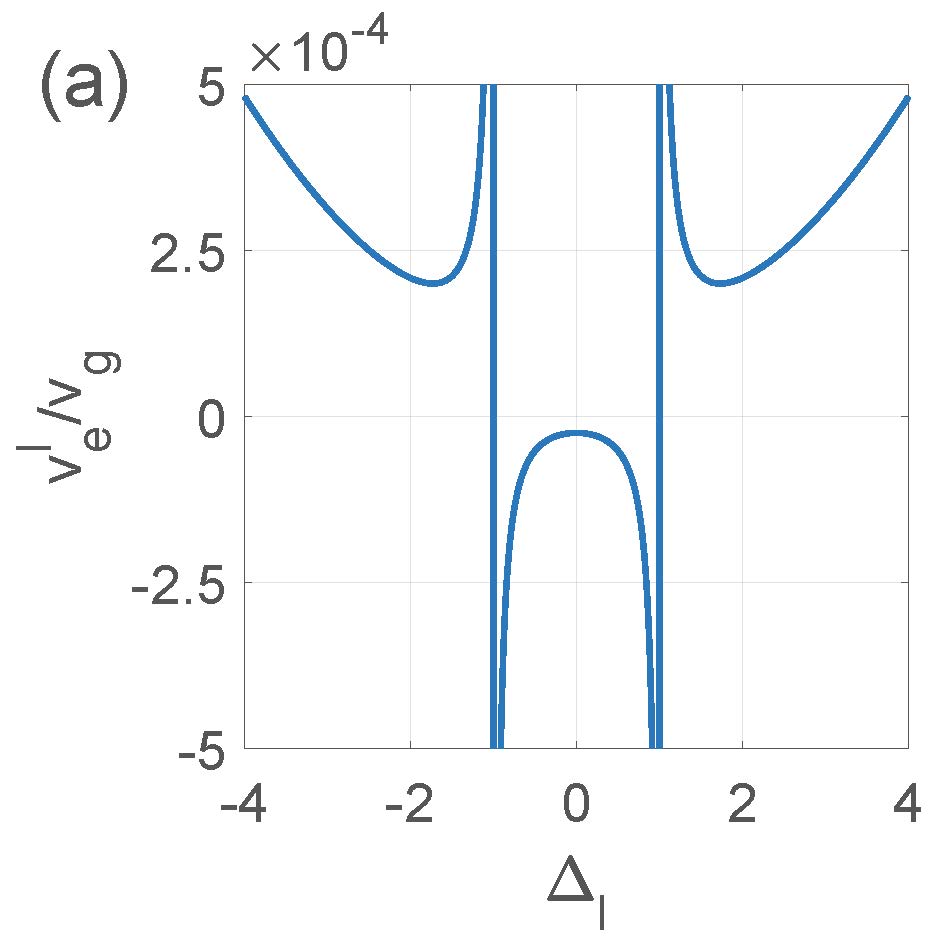}\includegraphics[width=0.5\linewidth]{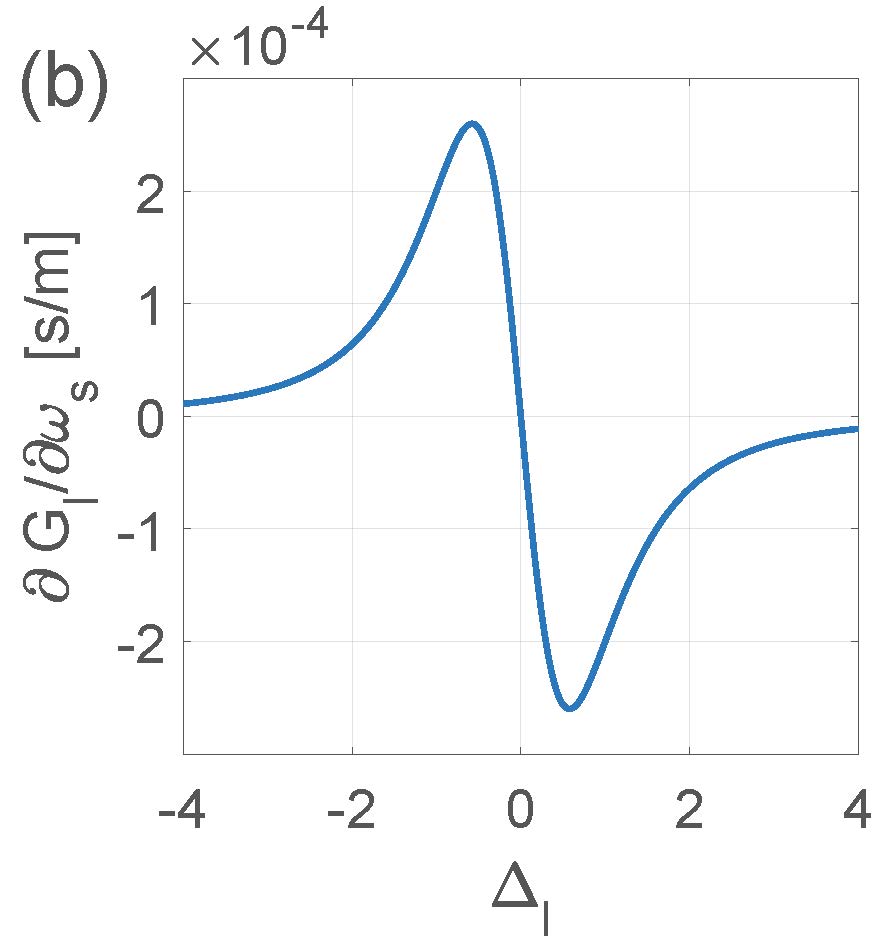}
\caption{(a) The relative effective group velocity $\frac{v_e^l}{v_g}$ as a function of the scaled detuning $\Delta_l$. (b) The rate of change of the gain factor with respect to the signal frequency $\frac{\partial G_l}{\partial\omega_s}$ as a function of the scaled detuning $\Delta_l$.}
 \label{Fig7}
\end{figure}

In figure (\ref{Fig4}), we plot the gain factor $G_lL$ from equation (\ref{Gl}) as a function of $\Delta_l$. In figure (\ref{Fig7}.a), the relative effective velocity $\frac{v_e^l}{v_g}$ from equation (\ref{vel}) is plotted as a function of $\Delta_l$. The rate of change of the gain factor with respect to the signal frequency, $\frac{\partial G_l}{\partial\omega_s}$ from equation (\ref{dGl}), is depicted in figure (\ref{Fig7}.b) as a function of $\Delta_l$. The plots demonstrate that the effective group velocity $v_e^l$ is significantly smaller than the group velocity $v_g$ around zero detuning. Meanwhile, the rate of change of the gain factor is negligible in the same region, but the loss factor $G_lL$ is substantial in this interval, leading to significant attenuation of the signal photons.

\section{Slow light without gain and loss}\label{Sec:SlowLightNoGain}

Based on the discussions, we conclude that achieving a slow signal within a waveguide while maintaining a constant signal amplitude using SBS with a single pump field is unattainable. Our primary interest lies in slowing down the signal field to the level of single photons. Our objective is to attain a propagating signal with an effective group velocity significantly lower than that in free space, while also ensuring a constant average number of quanta. Additionally, it's crucial to minimize the impact of thermal fluctuations, preventing them from significantly affecting the propagating signal. Therefore, our goal is to introduce a configuration that enables the realization of slow signals at the single-photon level without inducing gain or loss.

\begin{figure}[t]
\includegraphics[width=1\linewidth]{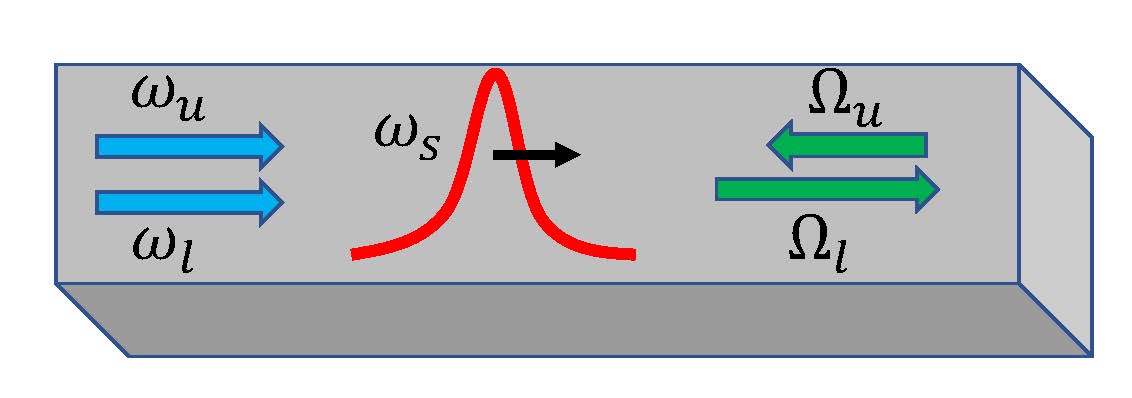}
\caption{A signal field of frequency $\omega_s$ is propagating to the right, with two co-propagating classical pump fields of frequencies $\omega_u$ and $\omega_l$, where $\omega_u>\omega_s>\omega_l$. Due to stimulated Brillouin scattering a signal photon scatters into a pump photon of frequency $\omega_l$ by the emission of a co-propagating phonon of frequency $\Omega$, and a pump photon of frequency $\omega_u$ scatters into a signal photon by the emission of a counter-propagating phonon of the same frequency.}
 \label{Fig8}
\end{figure}

\begin{figure}[t]
\includegraphics[width=1\linewidth]{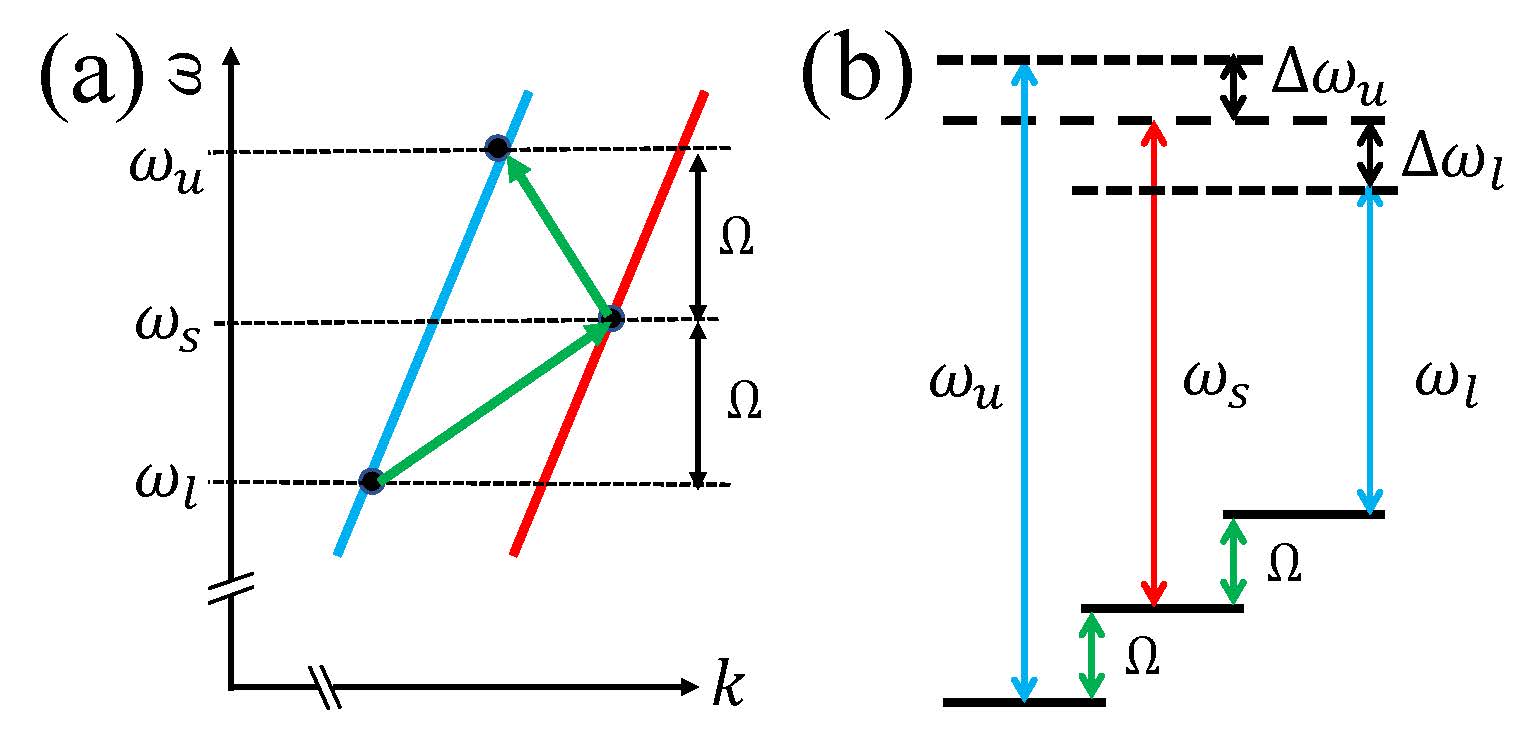}
\caption{(a) A pump field $(\omega_u)$ scatters into a signal field $(\omega_s)$ by the emission of a phonon $(\Omega)$, and a signal field scatters into a pump field $(\omega_l)$ by the emission of a phonon of the same frequency. The two phonon differs in their wavenumbers. (b) A schematic energy diagram of the photon and phonon modes for the two processes. A pump photon (of frequency $\omega_u$) is annihilated and a signal photon (of frequency $\omega_s$) and a phonon (of frequency $\Omega$) are created, with the detuning frequency $\Delta\omega_u=\omega_u-\omega_s-\Omega$. A signal photon is annihilated and a pump photon (of frequency $\omega_l$) and a phonon (of the same frequency) are created, with the detuning frequency $\Delta\omega_l=\omega_s-\omega_l-\Omega$.}
 \label{Fig9}
\end{figure}

To address the challenges previously discussed, we propose a unique configuration in which the signal field is coupled through SBS to two pump fields, involving a dispersion-less vibration mode. This approach aims to demonstrate that by merging the two aforementioned scenarios, a slow signal can be achieved without gain or loss, where the processes of signal amplification and attenuation counterbalance each other. Specifically, a signal with frequency $\omega_s$ and group velocity $v_g$ is coupled to two classical pump fields with amplitudes ${\cal E}_l$ and ${\cal E}_u$, and frequencies $\omega_l$ and $\omega_u$ respectively, where $\omega_u>\omega_s>\omega_l$, as depicted in figures (\ref{Fig8}) and (\ref{Fig9}). The involved dispersion-less vibration mode operates at frequency $\Omega$. The SBS process adheres to the phase matching condition for coupling with both the upper and lower pump fields. The photon-phonon coupling parameter is considered to be real, local (i.e., wavenumber independent), and identical for both interactions, with $g=g_l=g_u$. Additionally, the lower and upper detuning frequencies are defined as $\Delta\omega_l=\omega_s-\omega_l-\Omega$ and $\Delta\omega_u=\omega_u-\omega_s-\Omega$, respectively, as schematically illustrated in figure (\ref{Fig9}.b). Both the upper and lower SBS processes involve phonons at the same frequency $\Omega$ but with distinct wavenumbers. The phonon damping rate is denoted by $\Gamma$, and the Langevin force operator $\hat{\cal F}$ is considered identical for both Brillouin scattering processes.

The photon Hamiltonian is specified in (\ref{SigHam}), and the phonon Hamiltonian combines both upper and lower phonon contributions, $H_{\mathrm{phon}}=H_{\mathrm{phon}}^u+H_{\mathrm{phon}}^l$, utilizing Hamiltonians (\ref{phonu}) and (\ref{phonl}). Correspondingly, the photon-phonon interaction Hamiltonian merges the two interaction scenarios, $H_{\mathrm{phot-phon}}=H_{\mathrm{phot-phon}}^u+H_{\mathrm{phot-phon}}^l$, confer Equations (\ref{photphonu}) and (\ref{photphonl}).

In the interaction picture, the equation of motion for the photon operator is expressed as
\begin{multline}
\left(\frac{\partial}{\partial t}+v_g\frac{\partial}{\partial z}\right)\hat{\psi}_{s}(z,t)=-i\sqrt{L}g{\cal E}_u e^{-i\Delta\omega_ut}\ \hat{\cal Q}_{u}^{\dagger}(z,t)\\-i\sqrt{L}g{\cal E}_l e^{i\Delta\omega_lt}\ \hat{\cal Q}_{l}(z,t),
\end{multline}
and the phonon equations of motion follow from equations (\ref{eqphotphonu}) and (\ref{eqphotphonl}). Adopting a similar approach to that in appendices A and B for solving these equations, we arrive at
\begin{multline}
\hat{\psi}_{s}(z,t)=\hat{\psi}_{s}^{in}(z-v_gt)e^{(G-i\kappa)z} \nonumber \\
+i\frac{\sqrt{L}g}{v_g}\int_0^t dt'\int_0^{z}dz'e^{(G-i\kappa)(z-z')}e^{-\frac{\Gamma}{2}(t-t')}\\ \times\left\{{\cal E}_l\hat{\cal F}(z',t')e^{i\Delta\omega_lt}+{\cal E}_u\hat{\cal F}^{\dagger}(z',t')e^{-i\Delta\omega_ut}\right\}
\end{multline}
where $G=G_u-G_l$ and $\kappa=\kappa_u+\kappa_l$, integrating $G_u,\ \kappa_u$ from (\ref{Gu}, \ref{ku}), and $G_l,\ \kappa_l$ from (\ref{Gl}, \ref{kl}). The gain $G$ and phase shift $\kappa$ are given by
\begin{equation}
G=\frac{2g^2}{v_g\Gamma}\left\{\frac{{\cal I}_u}{1+\Delta_u^2}-\frac{{\cal I}_l}{1+\Delta_l^2}\right\},
\end{equation}
and
\begin{equation}
\kappa=\frac{2g^2}{v_g\Gamma}\left\{\frac{\Delta_u{\cal I}_u}{1+\Delta_u^2}+\frac{\Delta_l{\cal I}_l}{1+\Delta_l^2} \right\}.
\end{equation}
The key control parameters remain the scaled detunings $\Delta_u=2\Delta\omega_u/\Gamma$ and $\Delta_l=2\Delta\omega_l/\Gamma$, alongside the dimensionless pump intensities ${\cal I}_u=L|{\cal E}_u|^2$ and ${\cal I}_l=L|{\cal E}_l|^2$.

For the photon density, we obtain
\begin{equation}
\langle\hat{\psi}_{s}^{\dagger}(z,t)\hat{\psi}_{s}(z,t)\rangle=\langle\hat{\psi}_{s}^{in\dagger}(z-v_gt)\hat{\psi}_{s}^{in}(z-v_gt)\rangle e^{2Gz}+{\cal N}(z,t),
\end{equation}
where the thermal fluctuation contribution is given by
\begin{equation}
{\cal N}(z,t)=-\frac{g^2}{2Gv_g^2}\left\{{\cal I}_l\bar{n}+{\cal I}_u(\bar{n}+1)\right\}\left(1-e^{-\Gamma t}\right)\left(1-e^{2Gz}\right).
\end{equation}
Utilizing relations (\ref{LF}) for both the upper and lower processes, correlations among the Langevin force operators corresponding to the upper and lower processes are neglected.

The effective group velocity is defined by
\begin{equation}
\frac{1}{v_e}=\frac{1}{v_g}-\frac{\partial\kappa}{\partial\omega_s}.
\end{equation}
We have
\begin{equation}\label{ve}
\frac{v_e}{v_g}=\left(1+\frac{4g^2}{\Gamma^2}\left\{{\cal I}_u\frac{\left[1-\Delta_u^2\right]}{\left[1+\Delta_u^2\right]^2}-{\cal I}_l\frac{\left[1-\Delta_l^2\right]}{\left[1+\Delta_l^2\right]^2}\right\}\right)^{-1}.
\end{equation}
The rate of change of gain with respect to the signal frequency is expressed as
\begin{equation}\label{dG}
\frac{\partial G}{\partial\omega_s}=\frac{8g^2}{v_g\Gamma^2}\left\{{\cal I}_u\frac{\Delta_u}{\left[1+\Delta_u^2\right]^2}+{\cal I}_l\frac{\Delta_l}{\left[1+\Delta_l^2\right]^2}\right\},
\end{equation}
The objective is to achieve a slow propagating signal, where $\frac{v_e}{v_g}\ll1$. Additionally, it is essential for the signal to propagate without gain or loss along the wire, indicated by $GL\ll1$. Concurrently, we aim to minimize the influence of thermal fluctuations, ensuring that ${\cal N}_{l}\ll 1$. Our goal is to determine the conditions necessary to satisfy these three requirements.

We aim to achieve propagating light without gain or loss, which is possible when $G_u \approx G_l$, leading to $GL \approx 0$. This condition can be satisfied by ensuring that
\begin{equation}\label{Geq0}
\frac{{\cal I}_u}{{\cal I}_l}\approx\frac{1+\Delta_u^2}{1+\Delta_l^2}.
\end{equation}
Additionally, the thermal fluctuation contribution to the signal needs to be significantly less than one. At the waveguide output, at $z=L$, in the limit $GL \ll 1$, and under the condition $\Gamma L/v_g \ll 1$, the thermal contribution is given by
\begin{equation}
{\cal N}_{out}\approx\frac{g^2\Gamma L^2}{v_g^3}\left\{{\cal I}_l\bar{n}+{\cal I}_u(\bar{n}+1)\right\}.
\end{equation}
The contribution of thermal fluctuations to the average number of signal photons at the waveguide output should also be much smaller than one, i.e., ${\cal N}_{out} \ll 1$.

For further analysis of the result, we define the ratios $a = \frac{{\cal I}_u}{{\cal I}_l}$ and $b = \frac{\Delta_u}{\Delta_l}$. We use ${\cal I}_l = {\cal I}$, then ${\cal I}_u = a {\cal I}$, and $\Delta_l = \Delta$ then $\Delta_u = b \Delta$. The requirement (\ref{Geq0}) is written as $\Delta^2 = \frac{1-a}{a-b^2}$. Note that $1 < a < b^2$ or $b^2 < a < 1$. For example, we use the previous physical values, with ${\cal I} = 10^8$. We choose $a = b = \frac{1}{4}$ then $\Delta = 2$. We get $\frac{v_e}{v_g} \approx 10^{-4}$, and $\frac{\partial G}{\partial\omega_s} \approx 1.28 \times 10^{-4}$ s/m. We obtain a slow light with relatively large bandwidth without gain or loss. For the thermal contribution we get ${\cal N}_{out} \approx 2.8 \times 10^{-3}$.

\begin{figure}[t]
\includegraphics[width=0.52\linewidth]{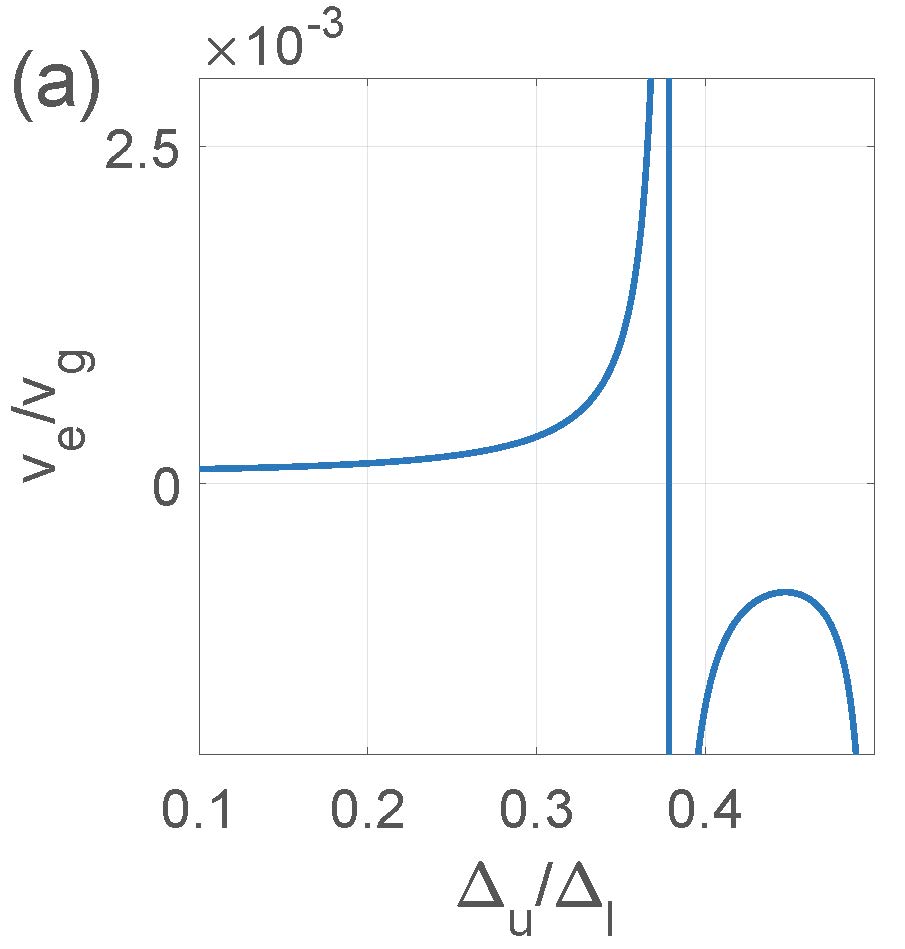}\includegraphics[width=0.5\linewidth]{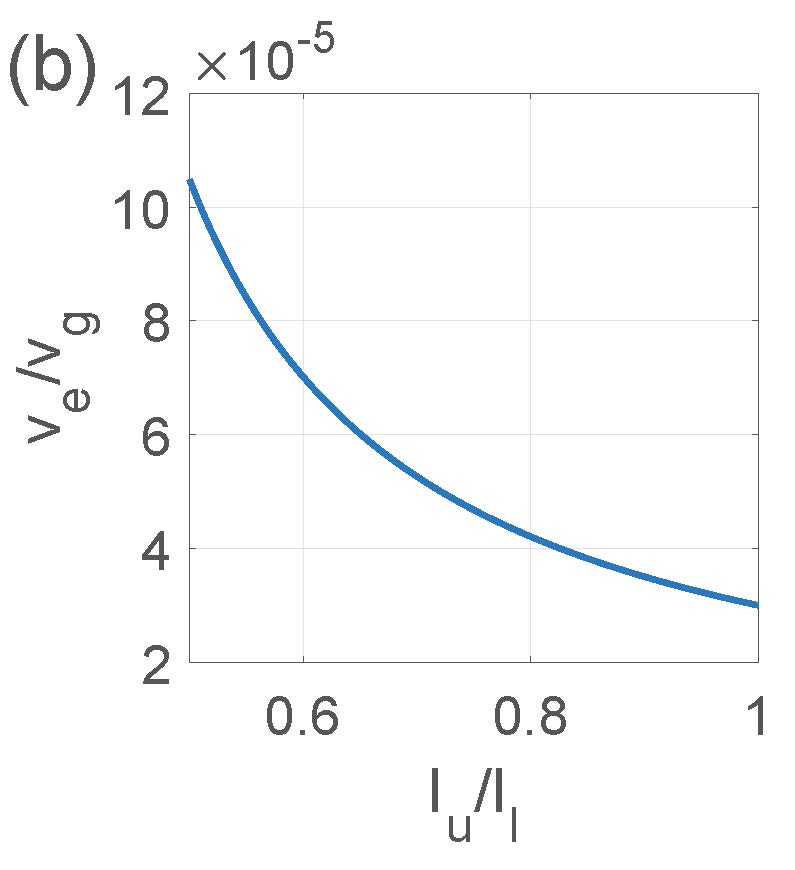}
\caption{(a) The relative effective group velocity $\frac{v_e}{v_g}$ as a function of the relative scaled detuning $\frac{\Delta_u}{\Delta_l}$. (b) The relative effective group velocity $\frac{v_e}{v_g}$ as a function of the relative pump intensity $\frac{{\cal I}_u}{{\cal I}_i}$.}
 \label{Fig10}
\end{figure}

\begin{figure}[t]
\includegraphics[width=0.53\linewidth]{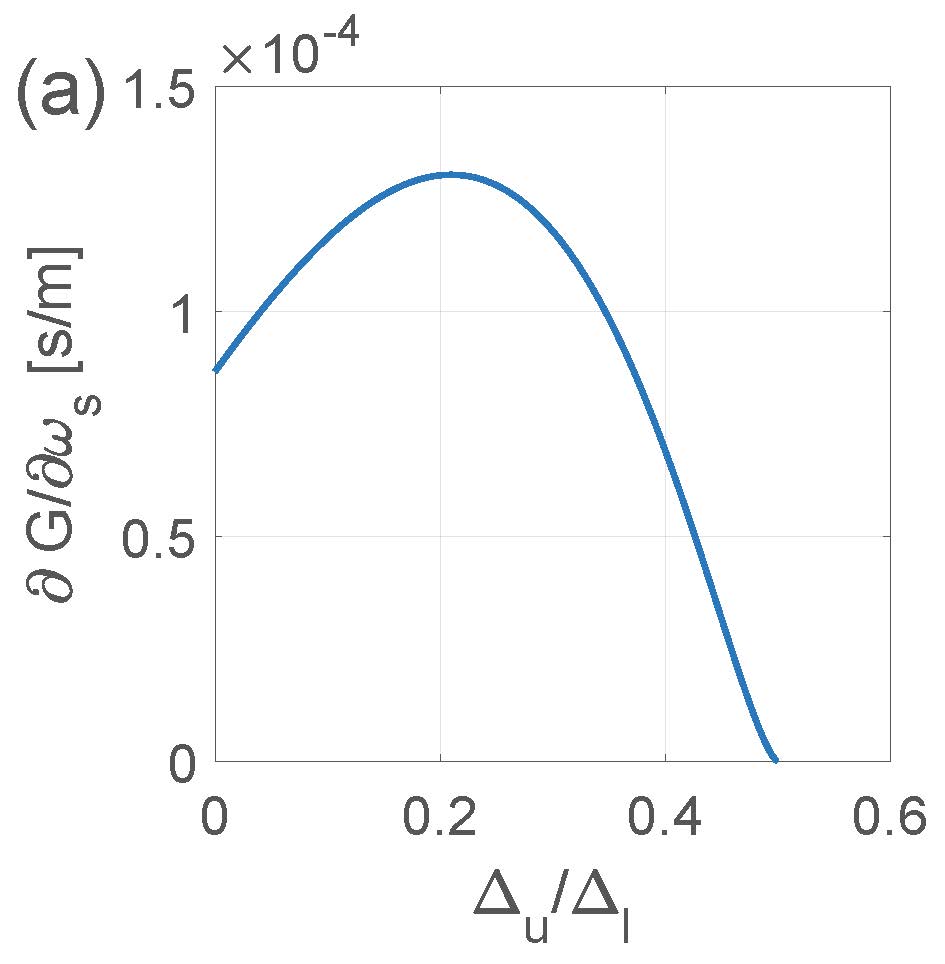}\includegraphics[width=0.5\linewidth]{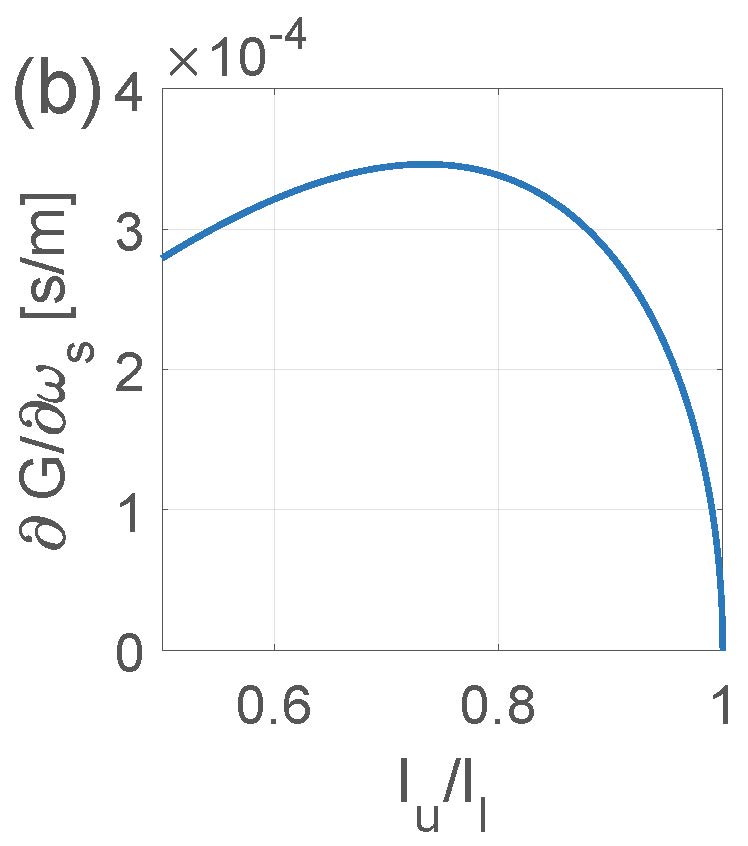}
\caption{(a) The rate of change of the gain factor with respect to the signal frequency $\frac{\partial G}{\partial\omega_s}$  as a function of the relative scaled detuning $\frac{\Delta_u}{\Delta_l}$. (b) The rate of change of the gain factor with respect to the signal frequency $\frac{\partial G}{\partial\omega_s}$  as a function of the relative pump intensity $\frac{{\cal I}_u}{{\cal I}_i}$.}
 \label{Fig11}
\end{figure}

For the case of zero gain $G=0$, the relative effective velocity $\frac{v_e}{v_g}$ from equation (\ref{ve}) is plotted in figure (\ref{Fig10}.a) as a function of $\Delta_u/\Delta_l$ for $I_u/I_l=1/4$, and in figure (\ref{Fig10}.b) as a function of $I_u/I_l$ for $\Delta_u/\Delta_l=1/4$. The rate of change of the gain factor with respect to the signal frequency $\frac{\partial G_l}{\partial\omega_s}$ from equation (\ref{dG}) is plotted in figure (\ref{Fig11}.a) as a function of $\Delta_u/\Delta_l$ for $I_u/I_l=1/4$, and in figure (\ref{Fig11}.b) as a function of $I_u/I_l$ for $\Delta_u/\Delta_l=1/4$. The effective group velocity $v_e$ is significantly smaller than the group velocity $v_g$, where $\frac{v_e}{v_g}\approx 10^{-4}$, for detunings up to $\Delta_u/\Delta_l < 1/3$. Note that the rate of change of the gain factor is negligible in the same zone, allowing the propagation of a wide-band signal without gain or loss and with negligible thermal contribution.

\section{Discussions and Conclusions}\label{Sec:Conclusion}

Optical quantum information processing is currently a leading candidate for the development of quantum computers. Generally, the components used in quantum information processing differ from those used in communication, which implies a need for interfaces between devices with varying physical properties. Such interfacing can significantly affect the coherence of quantum information. Nanophotonic structures involving photons can serve purposes in both quantum communication and quantum computing. This setup marks a crucial step toward an all-optical on-chip platform, using the same photons for quantum communication and computing, thereby avoiding decoherence effects associated with interfacing. Interactions among photons are critical for developing optical quantum logic gates. One of the primary obstacles to fabricating efficient photon-based quantum logic gates is the rapid propagation of optical fields within extensive nanophotonic structures. The high speed of light in these structures limits the accumulation of the nonlinear phases necessary for operating quantum logic gates.

In this paper, we introduce a configuration that enables slow signal propagation at the single-photon level by exploiting stimulated Brillouin scattering (SBS) within waveguides. The signal field can be significantly slowed via Brillouin scattering, which involves a classical pump field and propagating phonons. When the pump frequency exceeds that of the signal, it results in a substantial amplification of the signal amplitude; conversely, a pump frequency lower than that of the signal causes notable attenuation. To achieve a slow signal field without gain or loss, we propose a novel configuration that utilizes two pump fields with frequencies both above and below that of the signal. This arrangement allows the effects of amplification and attenuation to counterbalance each other, thus enabling the signal to propagate at a constant amplitude with an effective group velocity significantly reduced compared to that in free space. Additionally, this configuration can accommodate slow signals over wide bandwidths, extending up to tens of megahertz. We also consider the effects of thermal fluctuations by calculating the scattering of the pump fields off thermal phonons into and out of the signal field and establish conditions under which thermal contributions are negligible.

Slow light has been realized in a free-space medium containing an atomic ensemble \cite{Tey2008,Hammerer2010}. The control over light propagation in an optical medium can be achieved through Electromagnetic Induced Transparency (EIT), which enables the generation of both fast and slow light. In this process, coherent destructive interference prevents excitation within the optical medium \cite{Lukin2001,Fleischhauer2005,Chang2014}. EIT inherently satisfies the phase-matching requirement due to the presence of atomic components. To illustrate EIT, we examine a three-level atom configured in a lambda scheme with two lower metastable states, $|g\rangle$ and $|s\rangle$, and a higher excited state $|e\rangle$, where the transition between the lower states is dipole-forbidden. A probe field near resonance with the dipole-allowed transition $|g\rangle\leftrightarrow|e\rangle$ is affected by a strong control field close to resonance with the transition $|s\rangle\leftrightarrow|e\rangle$. The control field induces a superposition of the probe field and a coherent mix of the lower atomic states, mapping the photon onto a collective state of the atomic ensemble. This configuration creates a transparent window with an extremely narrow transparency band for the probe field in an otherwise opaque atomic medium, significantly reducing the probe field's effective group velocity. 

EIT has been demonstrated in cavity optomechanics via coupling between vibrational modes and photon modes through radiation pressure \cite{Safavi-Naeini2011}, where photons and phonons are localized within the resonator and phase-matching occurs naturally \cite{Weis2010}. Brillouin scattering induced transparency was shown by utilizing long-lived propagating light and phonons in a silica resonator under the required phase-matching conditions \cite{Kim2015}. Moreover, higher-order side-band induced transparency in optomechanical systems \cite{Xiong2012}, and optomechanical group delays in spinning resonator \cite{Zhang2024}, have been demonstrated. The approach introduced in the current paper allows for the propagation of signals across a broader bandwidth than achievable with the EIT scheme. Here, the phonon component serves a role analogous to the atomic component in EIT, ensuring phase-matching for the Brillouin scattering between the signal and pump fields.

The generation of slow photons is important for fundamental physics, e.g., for quantum nonlinear optics at the level of single photons, which rely on the derivation of effective photon-photon interactions \cite{Zoubi2017}. Additionally, the formation of photon bound states is explored \cite{Zoubi2021}. Slow photons in waveguides provide a test system for studying quantum phases of a gas of interacting photons. Moreover, slow photons have practical applications in nanophotonics for physical implementation in quantum information and quantum communication. The time delay achieved by slowing photons inside waveguides can serve as a memory device, a critical component for quantum computing with photons. A time delay on the order of microseconds can be achieved once the effective group velocity approaches the velocity of sound waves inside a waveguide.

\section*{Acknowledgment}

KH acknowledges support through Deutsche Forschungsgemeinschaft (DFG, German Research Foundation) through Project-ID 390837967
- EXC 2123.

\section*{Appendix A}

In this appendix we present the steps for solving the equations of motion (\ref{eqphotphonu}) for the case of slow light with signal amplification. We transform the equations into interaction picture with respect to the free Hamiltonian
\begin{align}
H_0&=\hbar\omega_{s}\int dz\ \hat{\psi}_{s}^{\dagger}(z)\hat{\psi}_{s}(z)\nonumber\\
&\quad+\hbar\omega_{u}\int dz\ \hat{\psi}_{u}^{\dagger}(z)\hat{\psi}_{u}(z)+\hbar\Omega\int dz\ \hat{\cal Q}_{u}^{\dagger}(z)\hat{\cal Q}_{u}(z),
\end{align}
with the replacement $\hat{\cal F}(z,t)\rightarrow e^{-i\Omega t}\hat{\cal F}(z,t)$. Note that the pump field is taken to be classical. The transformation is equivalent to the replacement $\hat{\psi}_{s}(z,t)\rightarrow e^{-i\omega_st}\hat{\psi}_{s}(z,t)$ and $\hat{\cal Q}_{u}(z,t)\rightarrow e^{-i\Omega t}\hat{\cal Q}_{u}(z,t)$ with ${\cal E}_u\rightarrow e^{-i\omega_ut}{\cal E}_u$. We obtain
\begin{align}
\left(\frac{\partial}{\partial t}+v_g\frac{\partial}{\partial z}\right)\hat{\psi}_{s}(z,t)&=-i\sqrt{L}g_{u}^{\ast}{\cal E}_u e^{-i\Delta\omega_ut}\ \hat{\cal Q}_{u}^{\dagger}(z,t), \nonumber \\
\left(\frac{\partial}{\partial t}+\frac{\Gamma}{2}\right)\hat{\cal Q}_{u}(z,t)&=-i\sqrt{L}g_{u}^{\ast}{\cal E}_u e^{-i\Delta\omega_ut}\ \hat{\psi}_{s}^{\dagger}(z,t)\nonumber\\
&\quad-\hat{\cal F}(z,t),
\end{align}
where we define the detuning frequency $\Delta\omega_u=\omega_u-\omega_s-\Omega$.

Formal integration of the phonon operator equation gives
\begin{align}
\hat{\cal Q}_{u}(z,t)&=\hat{\cal Q}_{u}(z,0)e^{-\Gamma t/2}\nonumber\\
&\quad-i\sqrt{L}g_{u}^{\ast}{\cal E}_u\int_0^tdt'\ \hat{\psi}_{s}^{\dagger}(z,t')\ e^{-i\Delta\omega_ut'}e^{-\frac{\Gamma}{2}(t-t')}\nonumber\\
&\quad-\int_0^tdt'\ \hat{\cal F}(z,t')e^{-\frac{\Gamma}{2}(t-t')}.
\end{align}
Iterative solution in term of the photon-phonon coupling parameter allows taking the signal operator out of the integral. In neglecting the phonon operator at initial time and after time integration, we get
\begin{align}
\hat{\cal Q}_{u}(z,t)&=-i\sqrt{L}g_{u}^{\ast}{\cal E}_u\frac{e^{-i\Delta\omega_ut}}{\Gamma/2-i\Delta\omega_u}\hat{\psi}_{s}^{\dagger}(z,t)\nonumber\\
&\quad-\int_0^tdt'\ \hat{\cal F}(z,t')e^{-\frac{\Gamma}{2}(t-t')}.
\end{align}
Substituting the phonon operator into the signal operator equation of motion gives
\begin{align}
\left(\frac{\partial}{\partial t}+v_g\frac{\partial}{\partial z}\right)\hat{\psi}_{s}(z,t)&=v_g(G_u-i\kappa_u)\hat{\psi}_{s}(z,t)\nonumber\\
&\quad+i\sqrt{L}g_{u}^{\ast}{\cal E}_u \hat{\cal W}_u^{\dagger}(z,t),
\end{align}
where the gain parameter is
\begin{equation}\label{Gu}
G_u=\frac{\Gamma|g_u|^2L|{\cal E}_u|^2}{2v_g\left(\Gamma^2/4+\Delta\omega_u^2\right)},
\end{equation}
and the wavenumber shift is
\begin{equation}\label{ku}
\kappa_u=\frac{\Delta\omega_u|g_u|^2L|{\cal E}_u|^2}{v_g\left(\Gamma^2/4+\Delta\omega_u^2\right)}.
\end{equation}
We defined
\begin{equation}\label{Wu}
\hat{\cal W}_{u}(z,t)=e^{i\Delta\omega_{u}t}\int_0^tdt'\ \hat{\cal F}(z,t')e^{-\frac{\Gamma}{2}(t-t')}.
\end{equation}

We apply the change of variables $\eta=z$ and $\xi=z-v_gt$, where $\frac{\partial}{\partial t}=-v_g\frac{\partial}{\partial\xi}$ and $\frac{\partial}{\partial z}=\frac{\partial}{\partial\eta}+\frac{\partial}{\partial\xi}$, then $\frac{\partial}{\partial t}+v_g\frac{\partial}{\partial z}=v_g\frac{\partial}{\partial\eta}$. The equation of motion in the new variables is
\begin{equation}
\frac{\partial}{\partial\eta}\hat{\psi}_{s}(\eta,\xi)=(G_u-i\kappa_u)\hat{\psi}_{s}(\eta,\xi)+i\frac{\sqrt{L}g_{u}^{\ast}{\cal E}_u}{v_g} \hat{\cal W}_u^{\dagger}(\eta,\xi).
\end{equation}
Formal integration leads to the solution
\begin{align}
\hat{\psi}_{s}(\eta,\xi)&=\hat{\psi}_{s}^{in}(\xi)e^{(G_u-i\kappa_u)\eta}\nonumber \\
&\quad+i\frac{\sqrt{L}g_{u}^{\ast}{\cal E}_u}{v_g}\int_0^{\eta}d\eta'\ \hat{\cal W}_u^{\dagger}(\eta',\xi)e^{(G_u-i\kappa_u)(\eta-\eta')},
\end{align}
where $\hat{\psi}_{s}^{in}(\xi)=\hat{\psi}_{s}(\eta=0,\xi)$.

\section*{Appendix B}

In this appendix we present the steps for solving the equations of motion (\ref{eqphotphonl}) for the case of slow light with signal attenuation. We transform the equations into the interaction picture, as before but with the replacement ${\cal E}_l\rightarrow e^{-i\omega_lt}{\cal E}_l$. We obtain
\begin{align}
\left(\frac{\partial}{\partial t}+v_g\frac{\partial}{\partial z}\right)\hat{\psi}_{s}(z,t)&=-i\sqrt{L}g_{l}^{\ast}{\cal E}_l e^{i\Delta\omega_lt}\ \hat{\cal Q}_{l}(z,t), \nonumber \\
\left(\frac{\partial}{\partial t}+\frac{\Gamma}{2}\right)\hat{\cal Q}_{l}(z,t)&=-i\sqrt{L}g_{l}{\cal E}_l^{\ast} e^{-i\Delta\omega_lt}\ \hat{\psi}_{s}(z,t)\nonumber\\
&\quad-\hat{\cal F}(z,t),
\end{align}
where the detuning frequency is $\Delta\omega_l=\omega_s-\omega_l-\Omega$.

Formal integration of the phonon operator equation gives
\begin{align}
\hat{\cal Q}_{l}(z,t)&=\hat{\cal Q}_{l}(z,0)e^{-\Gamma t/2}\nonumber\\
&\quad-i\sqrt{L}g_{l}{\cal E}_l^{\ast}\int_0^tdt'\ \hat{\psi}_{s}(z,t')\ e^{-i\Delta\omega_lt'}e^{-\frac{\Gamma}{2}(t-t')}\nonumber\\
&\quad-\int_0^tdt'\ \hat{\cal F}(z,t')e^{-\frac{\Gamma}{2}(t-t')}.
\end{align}
Iterative solution in term of the photon-phonon coupling parameter allows taking the signal operator out of the integral. In neglecting the phonon operator at initial time and after time integration, we get
\begin{align}
\hat{\cal Q}_{l}(z,t)&=-i\sqrt{L}g_{l}{\cal E}_l^{\ast}\frac{e^{-i\Delta\omega_lt}}{\Gamma/2-i\Delta\omega_l}\hat{\psi}_{s}(z,t)\nonumber\\
&\quad-\int_0^tdt'\ \hat{\cal F}(z,t')e^{-\frac{\Gamma}{2}(t-t')}.
\end{align}
Substituting the phonon operator into the signal operator equation of motion gives
\begin{align}
\left(\frac{\partial}{\partial t}+v_g\frac{\partial}{\partial z}\right)\hat{\psi}_{s}(z,t)&=-v_g(G_l+i\kappa_l)\hat{\psi}_{s}(z,t)\\
&\quad+i\sqrt{L}g_{l}^{\ast}{\cal E}_l \hat{\cal W}_l(z,t),
\end{align}
where the gain parameter
\begin{equation}\label{Gl}
G_l=\frac{\Gamma|g_l|^2L|{\cal E}_l|^2}{2v_g\left(\Gamma^2/4+\Delta\omega_l^2\right)},
\end{equation}
and the wavenumber shift is
\begin{equation}\label{kl}
\kappa_l=\frac{\Delta\omega_l|g_l|^2L|{\cal E}_l|^2}{v_g\left(\Gamma^2/4+\Delta\omega_l^2\right)}.
\end{equation}
We used the definition
\begin{equation}\label{Wl}
\hat{\cal W}_l(z,t)=e^{i\Delta\omega_lt}\int_0^tdt'\ \hat{\cal F}(z,t')e^{-\frac{\Gamma}{2}(t-t')}.
\end{equation}

We apply, as above, the change of variables $\eta=z$ and $\xi=z-v_gt$, hence the equation of motion in the new variables is
\begin{equation}
\frac{\partial}{\partial\eta}\hat{\psi}_{s}(\eta,\xi)=-(G_l+i\kappa_l)\hat{\psi}_{s}(\eta,\xi)+i\frac{\sqrt{L}g_{l}^{\ast}{\cal E}_l}{v_g} \hat{\cal W}_l(\eta,\xi).
\end{equation}
Formal integration leads to the solution
\begin{align}
\hat{\psi}_{s}(\eta,\xi)&=\hat{\psi}_{s}^{in}(\xi)e^{-(G_l+i\kappa_l)\eta}\nonumber\\&\quad+i\frac{\sqrt{L}g_{l}^{\ast}{\cal E}_l}{v_g}\int_0^{\eta}d\eta'\ \hat{\cal W}_l(\eta',\xi)e^{-(G_l+i\kappa_l)(\eta-\eta')},
\end{align}
where $\hat{\psi}_{s}^{in}(\xi)=\hat{\psi}_{s}(\eta=0,\xi)$. 

%\bibliography{CitationSL}

%

\end{document}